\documentclass[aps,prl,twocolumn,superscriptaddress,nofootinbib]{revtex4-2}

\usepackage{amsmath}
\usepackage{graphicx}
\usepackage{amssymb}
\usepackage{color}
\usepackage{mathrsfs}
\usepackage{verbatim}
\usepackage[colorlinks,linkcolor=blue,urlcolor=black,citecolor=blue,anchorcolor=blue]{hyperref}
\usepackage[T1]{fontenc}
\usepackage{orcidlink}
\usepackage{float}

\begin{document}


\title{Weak-Signal Amplification through Stochastic Resetting}

\author{Wei Wang~\orcidlink{0000-0002-1786-3932}}
\email{weiwangnuaa@gmail.com}
\email{These authors contributed equally to this work.}
\affiliation{School of Mathematics and Statistics, Northwestern Polytechnical University, Xi'an 710072, China}
\affiliation{Institute of Physics and Astronomy, University of Potsdam, 14476 Potsdam, Germany}

\author{Kaixuan Li~\orcidlink{0009-0000-7895-8532}}
\email{k.x.li@nuaa.edu.cn}
\email{These authors contributed equally to this work.}
\affiliation{State Key Laboratory of Mechanics and Control for Aerospace Structures, 
College of Aerospace Engineering, Nanjing University of Aeronautics and Astronautics, Nanjing 210016, China} \affiliation{Institute of Physics and Astronomy, University of Potsdam, 14476 Potsdam, Germany} 

\author{Jinjie Zhu~\orcidlink{0000-0001-7647-3806}}
\email{ jinjiezhu@nuaa.edu.cn}
\affiliation{State Key Laboratory of Mechanics and Control for Aerospace Structures, 
College of Aerospace Engineering, Nanjing University of Aeronautics and Astronautics, Nanjing 210016, China} 

\author{Andrey G. Cherstvy~\orcidlink{0000-0002-0516-9900}} 
\email{a.cherstvy@gmail.com} \affiliation{Institute of Physics and Astronomy, University of Potsdam, 14476 Potsdam, Germany} 

\author{Ralf Metzler~\orcidlink{0000-0002-6013-7020}}
\email{rmetzler@uni-potsdam.de}
\affiliation{Institute of Physics and Astronomy, University of Potsdam, 14476 Potsdam, Germany} 

\author{Yong Xu\,\orcidlink{0000-0002-8407-4650}}
\email{hsux3@nwpu.edu.cn}
\affiliation{School of Mathematics and Statistics, Northwestern Polytechnical University, Xi'an 710072, China}
\affiliation{MOE Key Laboratory for Complexity Science in Aerospace, Northwestern Polytechnical University, Xi'an 710072, China}

\begin{abstract}
Stochastic resetting (SR) offers a distinct route to
amplifying weak periodic signals in nonlinear systems. Here, we demonstrate SR-induced resonance (SRIR) in a system with double-well potential driven by a weak periodic force, where the amplitude of the mean periodic response exhibits a pronounced maximum at an optimal SR rate. The underlying mechanism exploits dynamical instability:
SR repeatedly brings the system close to an unstable
state, where the phase of the periodic force at each reset biases the subsequent motion
toward one of the stable states. The local instability then amplifies this weak bias into a large-scale
response. The SRIR arises from a competition between rare resets that limit phase-dependent
state selection and frequent resets that prevent the system from fully relaxing to
the selected state.
We develop a general renewal formulation for the periodic response and a coarse-grained approximation that
predicts the optimal resetting rate, revealing the
low-frequency scaling $r_{\rm opt}(\omega)\propto\omega^{2/3}$, in agreement with numerical
simulations.
SRIR is fundamentally distinct from conventional noise-induced resonance (NIR), which relies on noise-activated interwell transitions whose rates are periodically modulated by the external driving force. In the
weak-forcing and slow-driving regime, SRIR can achieve much stronger signal amplification than NIR.
SRIR persists for a finite spread of resetting positions, and our framework extends beyond bistability. We further discuss how SRIR and NIR can be selectively controlled. Our work proposes a novel approach to the classical paradigm of stochastic resonance.
\end{abstract}

\maketitle

\textit{Introduction.---}
Stochastic resetting (SR) provides a simple way of controlling nonequilibrium dynamics by intermittently returning a system to a prescribed state~\cite{EvansMajumdarSchehr2020,Pal2022,Gupta2022}. Such renewal events can fundamentally reshape nonequilibrium stationary
states and transient relaxation~\cite{EvansMajumdar2011,Majumdar2015,liang2025,sandev2022},
modify response and ergodic properties~\cite{Sokolov2023,sand2022,barkai2023,wang2022}, and strongly affect
first passage statistics, often generating an optimal finite SR
rate~\cite{Reuveni2016,ChechkinSokolov2018,pal2017,pal2019,DeBruyne2020,reuveni2014}. SR has also become experimentally accessible,
particularly in colloidal systems controlled by optical
tweezers~\cite{TalFriedman2020,Besga2020,Faisant2021}, with recent
extensions to active particle
systems~\cite{Vatash2025,Olsen2026,Goerlich2024}. The scope of SR extends further to stochastic thermodynamics
~\cite{Olsen2024,Gupta2020}, extreme-value statistics~\cite{meyl2015}, and
dynamics on complex networks~\cite{boyer2021}.
Despite these advances, how SR reshapes the response of nonlinear systems remains 
largely unexplored. This raises the central question of our work: Can SR itself induce resonant amplification of weak periodic signals?

Resonant amplification in nonlinear systems is conventionally associated
with stochastic resonance, also referred as noise-induced resonance (NIR), which arises from the interplay of
multistability, noise, and weak periodic signals
~\cite{McNamara1989,Wellens2004,Gammaitoni1998,Pikovsky2001}.
In a paradigmatic bistable system, the periodic forcing is on its own too weak to drive transitions between the two stable states, but it modulates the potential barriers, while noise enables barrier crossings and drives the transitions between the states.
NIR emerges when the noise-induced switching time becomes comparable to the periodic driving time, allowing transitions between the two stable states to follow periodic signal.
NIR was originally introduced to explain the
noise-assisted synchronization of glacial--interglacial climate
transitions with weak periodic orbital forcing
~\cite{Benzi1981,Benzi1982} and later
this noise-assisted amplification was demonstrated experimentally
in electronic bistable systems~\cite{FauveHeslot1983},
ring lasers~\cite{McNamaraRoy1988},
and tunnel diodes~\cite{MantegnaSpagnolo1994},
as well as in biological sensory systems
~\cite{Douglass1993,LevinMiller1996}. Noise can also optimize coherence in excitable systems without periodic signal,
giving rise to coherence resonance~\cite{Pikovsky1997,Gang1993,Longtin1997}. 

\begin{figure}
\centering
\includegraphics[width=1\linewidth]{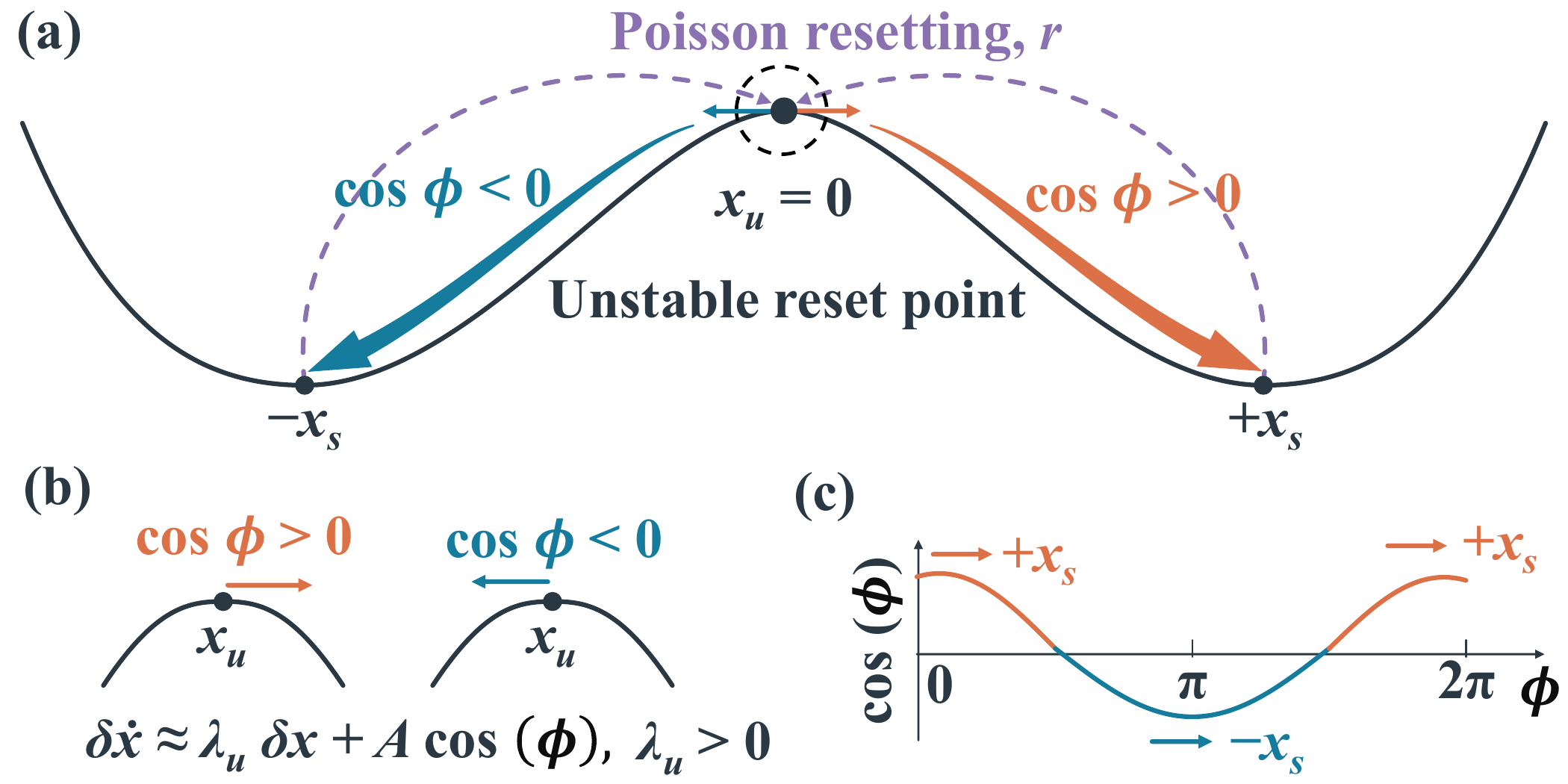}
\caption{
Schematic of SR-assisted amplification near an
unstable state in a bistable system driven by weak periodic force.
(a) Reset to instability: Poisson SR repeatedly returns
the system to the unstable state $x_r=0$,
where the instantaneous periodic bias favors relaxation
toward either $+x_s$ or $-x_s$.
(b) Weak periodic bias: Near the unstable state,
the periodic drive breaks the local symmetry throug
$\delta\dot{x}\approx\lambda_u\delta x+A\cos\phi$,
while the instability $\lambda_u>0$ amplifies
the resulting displacement.
(c) Phase-dependent selection: The phase at resetting
biases the relaxation direction to the stable states, with
$\cos\phi>0$ favoring $+x_s$ and
$\cos\phi<0$ favoring $-x_s$.
}
\label{fig:mechanism}
\end{figure}

SR provides a fundamentally different route, with reset events representing instantaneous state-reinitialization processes, ranging from abrupt climate transitions near tipping points \cite{Lenton2008}, voltage resets in bistable circuits, neuronal spike-and-reset dynamics \cite{Brette2005} to price dynamics of barrier-type options \cite{Vinod2022,Stojkoski2021}. SR repeatedly restores the system near a dynamically susceptible (typically unstable) state,
where the phase of the periodic drive at each reset determines the instantaneous directional bias and favors relaxation toward one of the stable states. The local instability then amplifies this weak phase-dependent bias into a large-scale response, see Fig.~\ref{fig:mechanism}.

In this Letter, we establish SR as a distinct route to resonant signal amplification in periodically driven systems, opening a new avenue for controlling response of nonlinear dynamics beyond conventional NIR. We demonstrate SR-induced resonance (SRIR) in a bistable system and uncover the physical picture: At small SR rates, rare resets provide too few opportunities for phase-dependent relaxation, whereas at large rates, frequent resets interrupt relaxation before the system can reach a selected stable state. This competition produces a pronounced maximum in the response amplitude at an intermediate SR rate. We derive an exact renewal formulation for the mean response under Poisson SR and obtain the corresponding optimality condition, while a coarse-grained approximation that predicts a robust scaling law for the optimal resetting rate.
Comparison with conventional NIR further shows that SRIR can achieve substantially stronger amplification of weak periodic signals over a broad range of driving frequencies.

\textit{Model.---}
We consider an overdamped particle in a double-well potential subject to periodic driving,
\begin{equation}
\frac{dx(t)}{dt}=ax(t)-bx^3(t)+A\cos(\omega t).
\label{eq:model}
\end{equation}
In addition, the particle undergoes Poisson SR at rate $r$. We denote the successive SR times by ${t_n}$, with ${n\geq1}$ and $t_0=0$, and define the waiting time between two successive SR events as $\tau_n=t_n-t_{n-1}$. The waiting times ${\tau_n}$ are independent and identically distributed according to the exponential density $\psi(\tau)=re^{-r\tau}$. At each SR time $t_n$, the particle is instantaneously returned to the reset point $x(t_n)=x_r=0$, while the phase of the periodic driving remains unchanged. In the absence of forcing, the potential $U(x)=-ax^2/2+bx^4/4$ with $a,b>0$, possesses two stable states at $x=\pm x_s$, with $x_s=\sqrt{a/b}$, separated by an unstable state $x_u=0$. We restrict our analysis to the subthreshold regime, $A<A_c=2a^{3/2}/(3\sqrt{3b})$, for which the periodic forcing alone cannot overcome the potential barrier \cite{footnote1}.

\textit{Renewal formulation.---}
Let $x_\phi(\tau)$ denote the uninterrupted deterministic trajectory following a reset that occurs at phase $\phi$. More precisely, if a reset occurs at time $t_n$, we define $\phi=\omega t_n (\mathrm{mod}~2\pi)$ and introduce the elapsed time $\tau=t-t_n$ measured from that reset event. The trajectory $x_\phi(\tau)$ then evolves from the initial condition $x_\phi(0)=x_r=0$ according to the deterministic dynamics \begin{eqnarray}
\frac{d{x}_\phi(\tau)}{d\tau}=ax_\phi(\tau)-bx_\phi^3(\tau)+A\cos(\phi+\omega\tau),  \label{eq:full_post_reset_dynamics}  
\end{eqnarray}provided no further reset occurs during the interval of duration $\tau$. We note that Eq.~(\ref{eq:model}) describes the full dynamics, including
a sequence of SR events, whereas Eq.~(\ref{eq:full_post_reset_dynamics}) describes a single
uninterrupted deterministic post-reset trajectory, parameterized by the
reset phase $\phi$ and the elapsed time $\tau$.
Representative post-reset trajectories generated by
Eq.~(\ref{eq:full_post_reset_dynamics}) are shown in
Fig.~\ref{fig:app-a}(a).
Near the unstable state $x_u=0$ and for sufficiently slow driving, the post-reset dynamics (\ref{eq:full_post_reset_dynamics}) can be
linearized as $dx_\phi/d\tau\sim ax_\phi+A\cos\phi$. This gives $x_\phi(\tau)\sim  \frac{A}{a}(e^{a\tau}-1)\cos\phi$. The sign of the instantaneous periodic bias at the reset time therefore defines the initial direction of motion, while the intrinsic instability of $x=0$ exponentially amplifies this small phase-dependent perturbation, driving the system toward a selected stable state.

As the most recent reset occurred at time $t_n=t-\tau$, the phase of the
periodic driving at that reset event is
$\phi=\omega(t-\tau)$, where the initial phase of the driving at $t=0$
is set to $\phi_0=0$ \cite{footnote2}.
The ensemble-averaged position then obeys the renewal
relation \cite{EvansMajumdarSchehr2020,ChechkinSokolov2018}
\begin{eqnarray}
    \left\langle x(t)\right\rangle
    =
    e^{-rt}x_{\phi_0}(t)
    +
    \int_0^t
    r e^{-r\tau}
    x_{\omega(t-\tau)}(\tau)
    d\tau .
    \label{eq:finite_renewal_mean}
\end{eqnarray}
Here, the first term represents the trajectories that experience no reset
before time $t$, whereas the integral accounts for
trajectories whose most recent reset occurred at time $t_n=t-\tau$.
Specifically, $x_{\phi_0}(t)$ denotes the deterministic trajectory evolving from the initial position with driving phase $\phi_0$, while $x_{\omega(t-\tau)}(\tau)$ denotes the deterministic trajectory evolving for a duration $\tau$ from the resetting position, with driving phase $\omega(t-\tau)$ at the reset time. The external driving phase remains unaffected by resetting.
Accordingly, $r e^{-r\tau}d\tau$ is the probability that the most
recent reset occurred within the interval
$(t-\tau-d\tau,t-\tau)$ and that no subsequent reset occurred prior to
time $t$.

In the long-time regime, $t\gg r^{-1}$, the no-reset contribution in
Eq.~(\ref{eq:finite_renewal_mean}) is exponentially suppressed.
As the deterministic post-reset trajectories remain bounded, the
upper integration limit can also be extended to infinity, yielding
$\langle x(t)\rangle=\int_0^\infty r e^{-r\tau}
x_{\omega(t-\tau)}(\tau)\,d\tau$, where the phase argument is understood
modulo $2\pi$. Expanding the $2\pi$-periodic post-reset trajectory as
$x_\phi(\tau)=\sum_n h_n(\tau)e^{in\phi}$, with the Fourier coefficients $h_n(\tau)$, gives
\begin{subequations} \label{renewal-response}
\begin{eqnarray}
    \left\langle x(t)\right\rangle
    &=&
    \sum_{n=-\infty}^{\infty}
    C_n(r)e^{in\omega t},
    \label{eq:response_fourier_a}\\
    C_n(r)
    &=&
    r\int_0^\infty
    e^{-(r+in\omega)\tau}
    h_n(\tau)\,d\tau ,    \label{eq:response_fourier_b}\\
    h_n(\tau)&=&(2\pi)^{-1}\int_0^{2\pi}
x_\phi(\tau)e^{-in\phi}\,d\phi.\label{eq:response_fourier_c}
\end{eqnarray}
\end{subequations}
The fundamental harmonic $n=1$ of the mean response has the amplitude $2|C_1|$. We therefore
define the amplification factor with respect to the driving-force magnitude $A$ as
\begin{eqnarray}
    G(r)
    &=&
    \frac{2|C_1(r)|}{A}
    \nonumber\\
    &=&
    \frac{r}{\pi A}
    \left|
    \int_0^{2\pi}d\phi
    \int_0^\infty d\tau\,
    x_\phi(\tau)
    e^{-r\tau}
    e^{-i(\phi+\omega\tau)}
    \right|.
    \label{eq:gain}
\end{eqnarray}
This exact representation retains the complete phase-dependent
post-reset dynamics of Eq.~(\ref{eq:full_post_reset_dynamics}) without
coarse graining. More generally, the same renewal structure applies to 
periodically driven nonlinear systems under Poisson
SR, with the model-specific dynamics entering through the
uninterrupted post-reset harmonics $h_n(\tau)$, see
Supplemental Material~\cite{SM}. 

At the maximum of $G(r)$, differentiation of
Eq.~(\ref{eq:gain}) with respect to $r$ gives the optimality
condition
\begin{equation}
    r_{\rm opt}\times
    \tau_{\rm eff}(r_{\rm opt},\omega)
    =
    1,
    \label{eq:matching}
\end{equation}
where 
\begin{equation}
    \tau_{\rm eff}(r,\omega)
    =
    {\rm Re}
    \left[
    \frac{
    \displaystyle
    \int_0^\infty
    \tau e^{-(r+i\omega)\tau}
    h_1(\tau)\,d\tau
    }{
    \displaystyle
    \int_0^\infty
    e^{-(r+i\omega)\tau}
    h_1(\tau)\,d\tau
    }
    \right].
    \label{eq:tau_eff}
\end{equation}
Equation~(\ref{eq:matching}) gives the exact optimality condition for
the resonant amplification, relating the optimal SR rate to a
weighted post-reset time scale. However, this condition does not necessarily imply the existence of a nonzero optimal SR rate in general periodically driven systems; an interior maximum requires additional structure in the system's response. A counterexample is a particle confined in a harmonic potential, $U(x)=kx^2/2$, which possesses a single stable equilibrium and exhibits maximum amplification at the boundary $r_{\rm opt}=0$, see Eqs.~(S24)–(S27) of Supplemental Material~\cite{SM}.

\begin{figure}[t]
\centering
\includegraphics[width=1\linewidth]{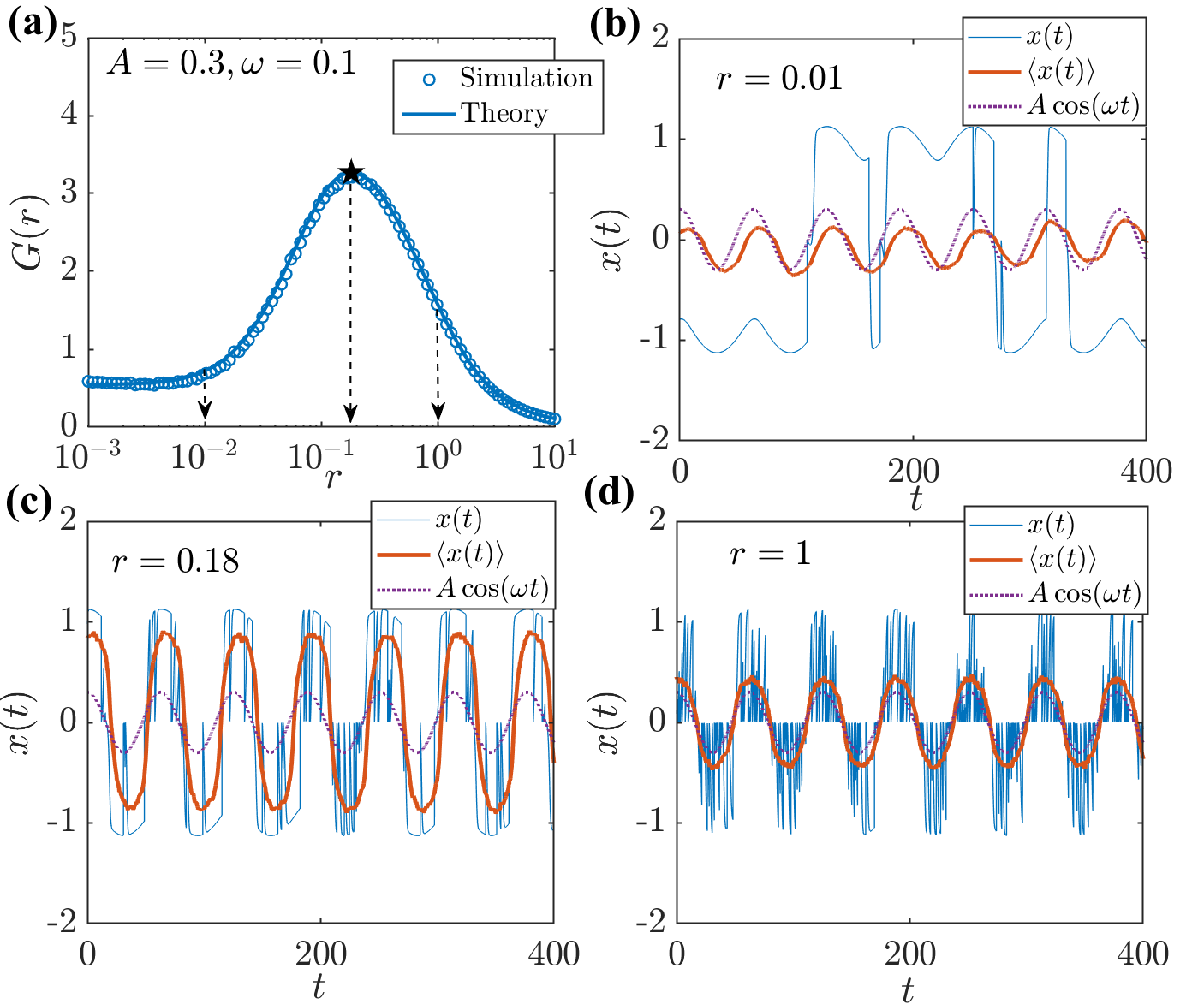}
\caption{
SRIR in the bistable system~(\ref{eq:model}).
(a) Amplification factor $G(r)$ for $A=0.3$ and
$\omega=0.1$. Symbols denote direct simulations,
and the solid line represents the exact renewal
prediction~(\ref{eq:gain}), yielding the maximum value at SR rate
$r_{\rm opt}\approx 0.18$.
The arrows mark the SR rates examined in
(b)--(d).
(b)--(d) Representative individual trajectories
$x(t)$ and ensemble-averaged responses
for $r=0.01$,
$r=r_{\rm opt}\approx 0.18$, and $r=1$, respectively.
The dotted curves are the periodic driving
$A\cos(\omega t)$.
Other parameters are $dt=0.01$, $N=500$, and
$a=b=x_s=1$ throughout the text.
}
\label{fig:resonance}
\end{figure}

To verify the exact renewal theory for the amplification factor (\ref{eq:gain}), we provide direct Monte Carlo simulations. Both approaches reveal a nonmonotonic dependence of the amplification on the SR rate in system (\ref{eq:model}), with excellent quantitative agreement [Fig.~\ref{fig:resonance}(a)]. Near the optimal rate, the ensemble-averaged response exhibits its largest first-harmonic amplitude [Fig.~\ref{fig:resonance}(c)], although individual trajectories remain stochastic due to a random sequence of resetting times.
The two limiting cases reveal the competing requirements underlying the
SRIR. For small $r$, resets are too rare to sustain 
phase-dependent selection despite allowing ample post-reset relaxation
[Fig.~\ref{fig:resonance}(b)]. For too large $r$, phase-selection events are
frequent, but repeated resets interrupt the full relaxation before a
selected stable state is reached [Fig.~\ref{fig:resonance}(d)]. The optimal SR rate therefore balances frequent phase selection against sufficient time for relaxation toward the selected stable state. In particular, for $r=0$, the amplification factor $G(r)$ approaches the finite value $1/\sqrt{4a^2+\omega^2}$, as predicted by linear response theory in the weak-driving regime~\cite{Gammaitoni1998}. The SR-induced dynamics of the bistable system (\ref{eq:model})
are further illustrated by Movie S1 in Supplemental Material~\cite{SM}.

\textit{Coarse-grained reduction.---}
Although the renewal formulation above is exact, the optimal SR
rate remains implicit because it depends on phase-dependent
post-reset trajectories. To obtain a transparent analytical description
and an explicit asymptotic prediction for $r_{\rm opt}$, we introduce a
coarse-grained representation of the post-reset dynamics.

In conventional NIR, a bistable system is commonly
described by a two-state reduction, in which the continuous dynamics is
coarse grained into occupations of the two wells and noise-activated
transitions between them~\cite{McNamara1989,Gammaitoni1998}. Such a
description is appropriate when the transition time is negligible
compared with the residence time in either well. In the present
SR problem, however, each reset returns the particle to the
unstable state, from which it undergoes a finite deterministic
relaxation before reaching a selected stable state. This transient
stage cannot be treated as instantaneous, because repeated SR
can interrupt the relaxation before the selected response is fully
established. Although the relaxation time may remain shorter than the
mean reset interval, it generates the leading response loss that
increases with the SR rate and therefore enters a balance
determining the optimum. We thus retain the finite post-reset relaxation
as an additional transient state in this simplified coarse-grained description.

Specifically, we introduce a three-state coarse-grained description
in which each exact post-reset trajectory is approximated as
\begin{equation}
x_\phi(\tau)
\sim
\sigma_\phi x_s\Theta[\tau-\tau_s(\phi)]
=
\begin{cases}
0, & 0\leq \tau<\tau_s(\phi)\\[1mm]
+x_s& \tau\geq\tau_s(\phi),\sigma_\phi=+1\\[1mm]
-x_s, & \tau\geq\tau_s(\phi),\sigma_\phi=-1
\end{cases},
\label{eq:3state}
\end{equation}
where $\Theta(z)=0$ for $z<0$ and
$\Theta(z)=1$ for $z\geq0$, is the Heaviside step function. $\tau_s(\phi)=\inf\{\tau>0:|x_\phi(\tau)|=x_s\}$ is the 
reset-to-well time, with $\tau_s(\phi)=\infty$ if the trajectory
never reaches either well. For $\tau_s(\phi)<\infty$,
$\sigma_\phi={\rm sgn}[x_\phi(\tau_s(\phi))]=\pm1$, is the index identifying the destination well. Both quantities are obtained
directly from the exact deterministic dynamics in
Eq.~(\ref{eq:full_post_reset_dynamics}) and are illustrated in
Fig.~\ref{fig:app-a}. Thus, the three coarse-grained states
correspond to a transient state $0$ and two well states
$\pm x_s$.

This reduction retains the phase-dependent transit time and potential-well
selection, $\{\tau_s(\phi),\sigma_\phi\}$, but neglects the
continuous position response during transit and the subsequent
intrawell motion. Substituting Eq.~(\ref{eq:3state}) into the renewal response (\ref{renewal-response}) yields
the approximate amplification factor and corresponding optimality condition,
Eqs.~(\ref{eq:EM_G3}) and (\ref{eq:EM_matching}), respectively,
as derived in End Matter.

For weak forcing and slow driving,
the reset-to-well relaxation time and selected-well index have
the asymptotic forms
\begin{equation}
    \tau_s(\phi)
    \sim
    \frac{3}{2a}
    \ln\!\left(\frac{a x_s}{A|\cos\phi|}\right),
    \qquad
    \sigma_\phi\sim{\rm sgn}(\cos\phi),
    \label{eq:weak_three_state}
\end{equation} as derived in Eqs.~(S49) and (S53) of the Supplemental
Material~\cite{SM}.
Substituting Eq.~(\ref{eq:weak_three_state}) directly into the reduced
amplification factor $G(r)$, Eq.~(\ref{eq:EM_G3}), 
and expanding in the asymptotic window $\overline{\tau}_s\ll r_{}^{-1}\ll\omega^{-1}$,
gives\begin{equation}
    G(r)
    \sim
    \frac{4x_s}{\pi A}
    \left(
        1-r\overline{\tau}_s-\frac{\omega^2}{2r^2}   
    \right),    \label{eq:three_state_gain_expansion}
\end{equation} where $
\overline{\tau}_s
=
\frac{3}{2a}
\left[
\ln\left(\frac{a x_s}{2A}\right)
+1
\right].$ The two corrections in Eq.~(\ref{eq:three_state_gain_expansion}) 
encode the competing response losses underlying
the SRIR. The term $\omega^2/(2r^2)$ results from insufficiently
frequent phase selection events and decreases as $r$ increases,
whereas $r\overline{\tau}_s$ describes the loss caused by resets that
interrupt the finite post-reset relaxation and increases with $r$.
Maximizing Eq.~(\ref{eq:three_state_gain_expansion}) with respect to
$r$ therefore yields
\begin{equation}
r_{\rm opt}
\sim
\left(
\frac{\omega^2}{\overline{\tau}_s}
\right)^{1/3}.
\label{eq:scaling}
\end{equation}
This predicts
$r_{\rm opt}(\omega)\propto\omega^{2/3}$, together with a much weaker
dependence on the forcing amplitude $A$ via $\overline{\tau}_s(A)$.

\begin{figure}[h]
    \centering
\includegraphics[width=0.7\linewidth]{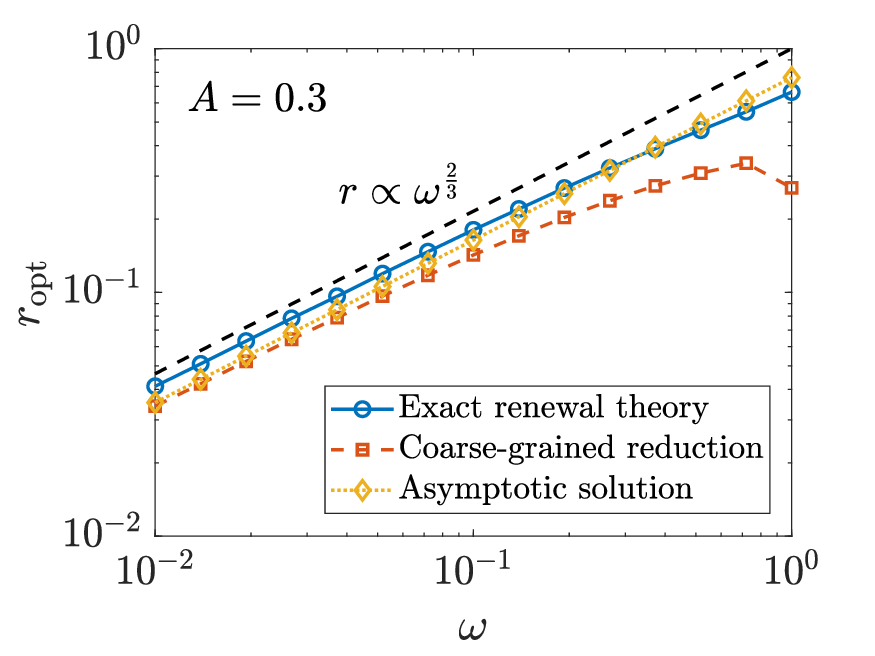}
    \caption{
Optimal SR rate $r_{\rm opt}$ versus driving frequency $\omega$
for $A=0.3$. Circles, squares, and diamonds denote the exact renewal
prediction~(\ref{eq:matching}), the coarse-grained prediction~~(\ref{eq:EM_matching}), and the weak-forcing,
slow-driving asymptotic result~(\ref{eq:scaling}), respectively.
}
\label{figure3}
\end{figure}

Figure~\ref{figure3} shows the dependence of the optimal SR rate
$r_{\rm opt}$ on the driving frequency $\omega$. The coarse-grained
reduction captures the increase of $r_{\rm opt}$ predicted by the exact
renewal theory (\ref{eq:matching}), with quantitative deviations appearing at higher
frequencies. In the low-frequency regime $\omega< \bar{\tau}_s^{-1}\approx0.44$ ($A=0.3, a=x_s=1$ in Fig.~\ref{figure3}), both results approach the
asymptotic prediction (\ref{eq:scaling}). Physically,
faster driving requires more frequent SR to maintain effective
phase selection, thereby shifting the resonance toward larger SR
rates. The extension of this asymptotic scaling to a broader class of
phase-selected responses beyond bistability, including systems with
three or four stable states~\cite{zhu2024}, is developed in Sec.~S2 of Supplemental Material~\cite{SM}.

\textit{Comparison with conventional NIR.---}To compare SRIR with conventional
NIR, we consider the same periodically
driven bistable system without SR, $r=0$, but subject to additive
Gaussian white noise $F=\sqrt{2D}\,\xi(t)$ on the right-hand side of Eq.~(\ref{eq:model}),
with
$\langle\xi(t)\rangle=0$ and
$\langle\xi(t)\xi(t')\rangle=\delta(t-t')$.
For each pair $(A,\omega)$, we solve the corresponding full-time periodic Fokker--Planck
equation \cite{Risken1989} and optimize the
resonant amplification factor $G_{\rm noise}(D)$ over the noise intensity $D$,
yielding
$G_{\rm noise}^{\max}=\max_D G_{\rm noise}(D)$, see the numerical details in Sec.~S4 of Supplemental Material~\cite{SM}. The noise-driven case does not rely on a Kramers-rate
or linear response approximation ~\cite{Kramers1940,Hanggi1990}.
We compare this directly with
$G_{\rm reset}^{\max}=\max_r G_{\rm reset}(r)$.
To provide a common reference for the two mechanisms, we also consider the
limiting case of fully phase-locked switching. For an ideal two-state
response satisfying $|\langle x(t)\rangle|\leq x_s$, the largest
possible first-harmonic amplitude is attained by a square wave that alternates between the two stable states every half-period of the weak drive, yielding
\begin{equation}
    G\leq G_{\rm sq}=\frac{4x_s}{\pi A}.
    \label{eq:square}
\end{equation}
The derivation is given in End Matter. This bound provides a natural
reference for how closely the optimized SRIR and NIR responses approach
the ideal square-wave limit.

\begin{figure}[t]
\centering
\includegraphics[width=1\linewidth]{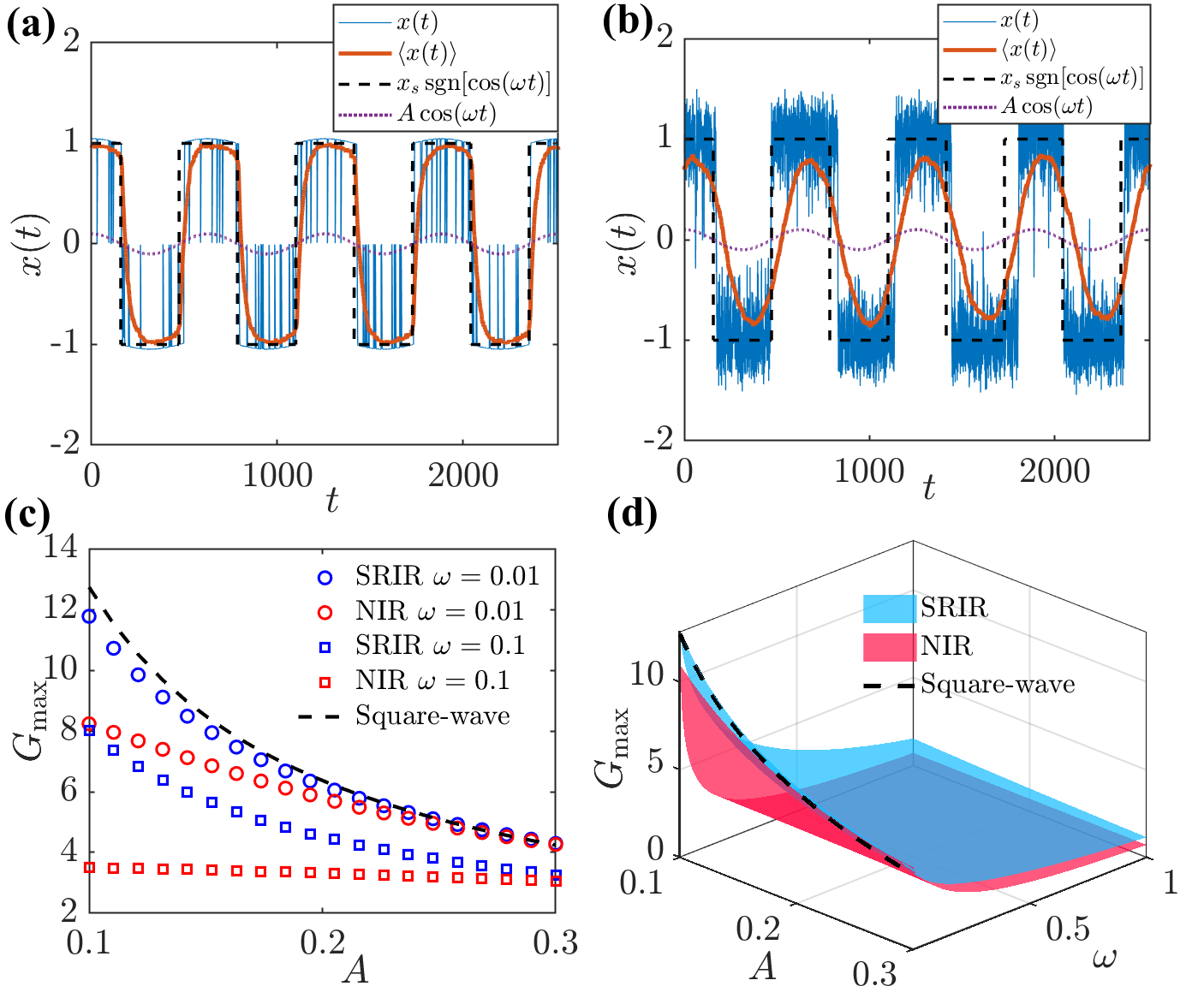}
\caption{
SRIR versus NIR. (a),(b) Representative dynamics at $A=0.1$ and $\omega=0.01$ for optimal SR ($r_{\rm opt}\approx0.035$) and optimal noise ($D_{\rm opt}\approx0.072$), respectively. Thin curves show individual trajectories, thick curves the ensemble-averaged responses, black dashed curves the ideal square-wave response, and dotted curves the weak periodic input $A\cos(\omega t)$. (c) Optimized amplification factors $G_{\rm reset}^{\max}=\max_r G_{\rm reset}(r)$ (blue) and $G_{\rm noise}^{\max}=\max_D G_{\rm noise}(D)$ (red) as functions of the forcing amplitude $A$ at $\omega=0.01$ (circles) and $0.1$ (squares). The black dashed curve denotes the ideal square-wave response (\ref{eq:square}). (d) Optimized amplification factors in the $(A,\omega)$ plane. The black dashed curve indicates the ideal square-wave response.
}
\label{fig:comparison}
\end{figure}

For the system with same double-well potential and periodic driving, at their respective optimal parameters, SRIR produces an
ensemble-averaged response much closer to square-wave response, whereas the NIR response remains less coherent, see Figs.~\ref{fig:comparison}(a) and \ref{fig:comparison}(b).
SR therefore approaches the square-wave limit more readily at
finite frequencies because it bypasses activated barrier crossing and
repeatedly returns the system directly to the unstable state.

Figure~\ref{fig:comparison}(c) further reveals quantitative differences between the two mechanisms and the ideal square-wave response. At $\omega=0.1$, the optimized noise-induced amplification approaches an approximately $A$-independent plateau, consistent with the linear-response prediction $G_{\rm noise}^{\max}\approx {\rm const}$ \cite{Gammaitoni1998}. By contrast, the optimized SR amplification factor $G_{\rm reset}^{\max}$ increases as the forcing amplitude $A$ decreases, approaching $\propto A^{-1}$ scaling of the ideal square-wave response~(\ref{eq:square}). Although both SRIR and NIR approach this reference in the limit $\omega\to0$, their finite-frequency behavior differs markedly. As shown in Fig.~\ref{fig:comparison}(d), optimized SR yields substantially stronger resonant amplification over a broad range of finite driving frequencies.

\textit{Potential experimental realization and applications.---}
The proposed SRIR could be accessible with
existing experimental ingredients. Periodically driven bistable
electronic circuits are well established~\cite{Gammaitoni1989}, and
Poisson-distributed control pulses could be used to reset the circuit
voltage to the vicinity of an unstable state. Similarly, optical
double-well platforms used to study stochastic
resonance~\cite{Jop2008} could be combined with holographic-tweezer
protocols for SR~\cite{TalFriedman2020}.
A further  platform is spin-Hall-effect
nanomagnets, where current pulses have been demonstrated
to prepare magnetization near a metastable 
configuration, from which a weak magnetic bias can
select a subsequent stable state~\cite{Bhowmik2014}.
In these realizations, instantaneous SR can be approached when the reset duration
is much shorter than the mean reset time,
the driving timescale, and the characteristic
relaxation time.

\textit{Discussion.---}
SRIR was also reported in generalized
Langevin dynamics with SR of the memory kernel
~\cite{Jolakoski2025}.
In our scenario here the mechanism is qualitatively different: SR acts directly on the dynamical state. This state-space reinitialization
provides a direct trajectory-level picture and lends itself naturally to
controlled experimental SR.
The
SRIR and signal amplification mechanism studied here are robust for a finite spread of SR positions
around the unstable state, see Eq.~(\ref{eq:robust_reset}) and Fig.~\ref{figure6} in End Matter ,
showing that precise resetting to the unstable equilibrium is not essential.

The results here are not restricted to the quartic double-well potential
or even to bistable dynamics. The renewal representation
(\ref{renewal-response}) and its coherent harmonics apply to a general
periodically driven system under independent Poisson SR.
The coarse-grained reduction can likewise be extended
to multi-attractor systems, defining a broader class of
phase-selected, finite-relaxation responses for which
$r_{\rm opt}(\omega)\propto\omega^{2/3}$ is valid.  A detailed discussion of the generality and possible extensions is provided in Supplemental Material~\cite{SM}, while the corresponding SRIR in a tristable system is illustrated in Supplemental Movie S2.
 
In the bistable
system (\ref{eq:model}) considered here, the periodic forcing may couple to both barrier
modulation and post-reset stable state selection, so SRIR and NIR can
coexist. An interesting question is whether SRIR and NIR can be selectively
controlled, so that one resonance remains active while the other is
suppressed. This possibility follows from a simple physical distinction
between their symmetry-breaking mechanisms: In conventional NIR, the periodic driving breaks the left-right symmetry by
alternately modulating the activation barriers, whereas in SRIR the driving breaks the symmetry of the post-reset dynamics.
In Sec.~S3 of Supplemental Material~\cite{SM}, we provide protocols to control the symmetry-breaking between  barrier modulation and
post-reset state selection,
in which the SRIR remains active while the NIR is suppressed, and vice versa.

Future studies may investigate the interplay between SR and noise, as well as extensions to non-Poisson resetting, partial resetting \cite{DiBello2023}, and optimized resetting protocols~\cite{pal2017,Bodrova2019}. 
Extending the present mechanism to underdamped dynamics ~\cite{Gupta2019,Franke2026} may lead to qualitatively different resonance behavior depending on the resetting protocol. While resetting both position and velocity preserves the renewal structure, resetting position alone retains velocity memory, potentially modifying the  amplification and the optimal resetting rate.

\textit{Acknowledgments.---} K.L. acknowledges funding by the China Scholarship Council (CSC Project 202406830056). J.Z. acknowledges the support from the National Natural Science Foundation of China (Grant No. 12572037). R.M. acknowledges financial support from the German Science Foundation (DFG, Grant No. 318763901, CRC Data Assimilation and ME 1535/22-1).

\textit{Data availability.---}
The code and data used to generate the simulation results and figures are available at \cite{code}.

\clearpage
\appendix

\onecolumngrid

\begin{center}
\textbf{END MATTER}
\end{center}

\twocolumngrid

\renewcommand{\theequation}{A\arabic{equation}}
\renewcommand{\theHequation}{A.\arabic{equation}}
\setcounter{equation}{0}

\renewcommand{\thefigure}{A\arabic{figure}}
\renewcommand{\theHfigure}{A.\arabic{figure}}
\setcounter{figure}{0}

\begin{figure}[h]
    \centering    \includegraphics[width=\linewidth]{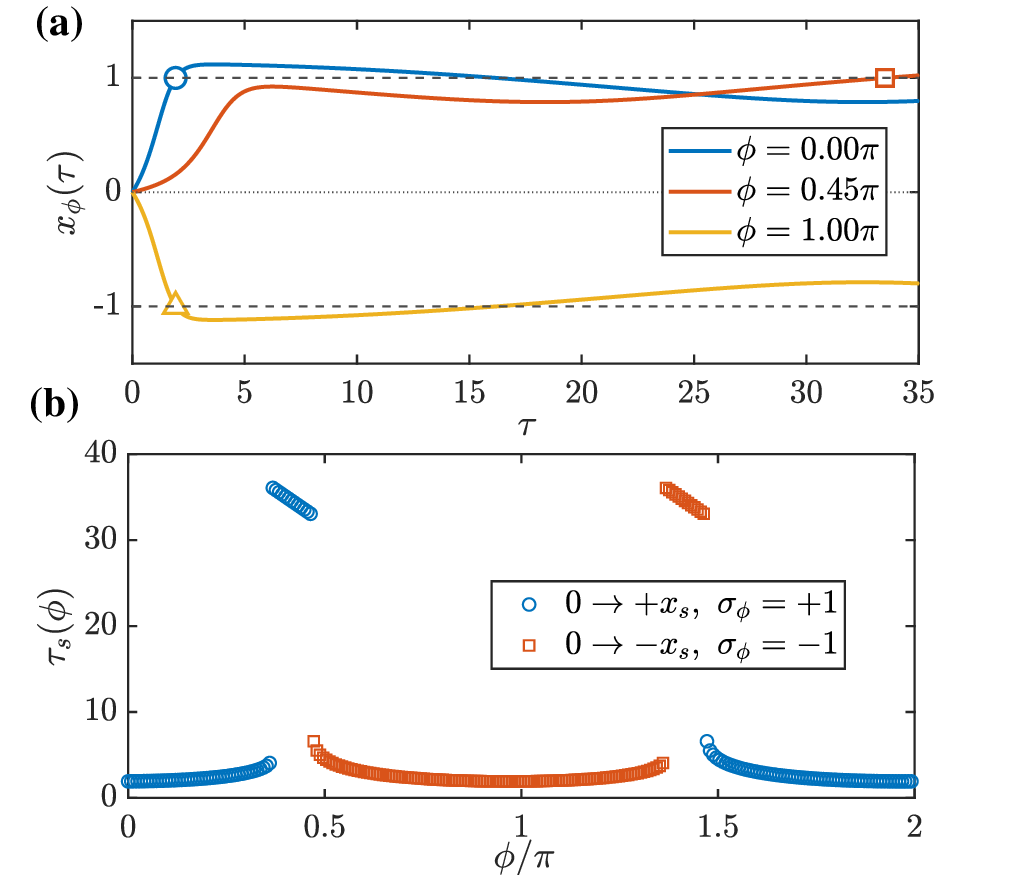}
\caption{
Phase-dependent reset-to-well dynamics underlying the coarse-grained reduction.
(a) Representative exact post-reset trajectories $x_\phi(\tau)$ for
reset phases $\phi$, obtained from Eq.~(\ref{eq:full_post_reset_dynamics}). Each trajectory
starts from $x_\phi(0)=0$ and subsequently follows the uninterrupted
deterministic dynamics. Open symbols mark the first arrival at either
$+x_s$ or $-x_s$, thereby defining the reset-to-well time
$\tau_s(\phi)$ and the destination variable $\sigma_\phi=\pm1$ used in
the coarse-grained reduction~(\ref{eq:3state}).
(b) Reset-to-well time $\tau_s(\phi)$ as a function of the reset
phase $\phi$, with the destination variable $\sigma_\phi$ encoded by
color: blue ($\sigma_\phi=+1$) and red ($\sigma_\phi=-1$) correspond
to first arrival at $+x_s$ and $-x_s$, respectively. Thus, panel (b)
displays the two phase-dependent quantities
$\{\tau_s(\phi),\sigma_\phi\}$ entering the coarse-grained reduction description (\ref{eq:3state}).
Parameters are $A=0.3$, $\omega=0.1$, and $x_s=1$.
}
\label{fig:app-a}
\end{figure}

\textit{Coarse-grained reduction of the exact renewal response.---}To establish a clear connection with the exact renewal formulation, we
substitute the coarse-grained post-reset trajectory representation~(\ref{eq:3state}) into the fundamental harmonic (\ref{eq:response_fourier_c}), yielding
\begin{equation}
    h_1(\tau)
    =
    \frac{x_s}{2\pi}
    \int_0^{2\pi}
    \sigma_\phi e^{-i\phi}
    \Theta[\tau-\tau_s(\phi)]\,d\phi .
    \label{eq:EM_h1}
\end{equation}
Here, $\tau_s(\phi)=\inf\{\tau>0:|x_\phi(\tau)|=x_s\}$ and
$\sigma_\phi={\rm sgn}[x_\phi(\tau_s(\phi))]$ are extracted directly from the
exact post-reset dynamics (\ref{eq:full_post_reset_dynamics}), as illustrated in
Fig.~\ref{fig:app-a}. The discontinuities in $\tau_s(\phi)$ shown in
Fig.~\ref{fig:app-a}(b) have two distinct origins.
Those accompanied by a change of $\sigma_\phi$ correspond to genuine
phase-selection boundaries. The remaining jumps do not indicate a
change of the selected well; they arise when the first relaxation
excursion narrowly misses the prescribed position $\pm x_s$, delaying
the first arrival until a later excursion. 

Inserting Eq.~(\ref{eq:EM_h1}) into the exact renewal coefficient~(\ref{eq:response_fourier_b}) and performing the
$\tau$ integration yields
\begin{equation}
    C_1(r)
    =
    \frac{x_s r}
    {2\pi(r+i\omega)}
    \int_0^{2\pi}
    \sigma_\phi e^{-i\phi}
    e^{-(r+i\omega)\tau_s(\phi)}
    d\phi,
    \label{eq:EM_C1}
\end{equation}
and the corresponding resonant amplification factor is
\begin{equation}
    G(r)
    =
    \frac{x_s r}
    {\pi A\sqrt{r^2+\omega^2}}
    \left|
    \int_0^{2\pi}
    \sigma_\phi e^{-i\phi}
    e^{-(r+i\omega)\tau_s(\phi)}
    d\phi
    \right|.
    \label{eq:EM_G3}
\end{equation}
The factors entering Eq.~(\ref{eq:EM_G3}) have a direct physical
interpretation: $\sigma_\phi$ retains the phase-selected destination
well, $e^{-r\tau_s(\phi)}$ is the probability that the reset-to-well
relaxation is completed without an intervening reset, and
$e^{-i\omega\tau_s(\phi)}$ accounts for the phase accumulated during
this relaxation.

Optimization of Eq.~(\ref{eq:EM_G3}) gives the optimality condition in coarse-grained description
\begin{equation}
    \frac{\omega^2}
    {r_{\rm opt}(r_{\rm opt}^2+\omega^2)}
    =
    \tau_{\rm eff}
    (r_{\rm opt},\omega),
    \label{eq:EM_matching}
\end{equation}
where the weighted reset-to-well time is
\begin{equation}
    \tau_{\rm eff}(r,\omega)
    =
    {\rm Re}
    \left[
    \frac{
    \displaystyle
    \int_0^{2\pi}
    \tau_s(\phi)\sigma_\phi e^{-i\phi}
    e^{-(r+i\omega)\tau_s(\phi)}
    d\phi
    }{
    \displaystyle
    \int_0^{2\pi}
    \sigma_\phi e^{-i\phi}
    e^{-(r+i\omega)\tau_s(\phi)}
    d\phi
    }
    \right].
    \label{eq:EM_taueff}
\end{equation}
This relation retains the same competition between SR and
post-reset relaxation as the exact renewal condition~(\ref{eq:matching}),
but is modified by the step-like coarse graining of the post-reset
trajectory. The exact renewal condition (\ref{eq:matching}) provides the quantitative prediction for
$r_{\rm opt}$, whereas the coarse-grained reduction (\ref{eq:EM_matching}) expresses the
resonance directly in terms of the phase dependent reset-to-well time
and enables analytical  asymptotic scaling (\ref{eq:scaling}).

\renewcommand{\theequation}{B\arabic{equation}}
\renewcommand{\theHequation}{B.\arabic{equation}}
\setcounter{equation}{0}

\renewcommand{\thefigure}{B\arabic{figure}}
\renewcommand{\theHfigure}{B.\arabic{figure}}
\setcounter{figure}{0}

\textit{Ideal two-state square-wave response.---}We derive the maximal fundamental harmonic response attainable
within an ideal two-state switching and clarify the square-wave
response used in the main text. Consider a periodic ensemble-averaged
response $\langle x(t)\rangle$ with period $T=2\pi/\omega$.
In the ideal two-state switching limit, the system occupies only the two stable
states $\pm x_s$, and therefore
\begin{equation}
    |\langle x(t)\rangle|
    \le
    x_s.
    \label{eq:EM_two_state_bound}
\end{equation}
The coherent first harmonic amplitude can be written as
\begin{equation}
    X_\omega
    =
    \max_{\varphi}
    \frac{2}{T}
    \int_0^T
    \langle x(t)\rangle
    \cos(\omega t-\varphi)\,dt ,
    \label{eq:EM_Xomega_phase}
\end{equation}
where $\varphi$ accounts for the phase lag of the response relative
to the periodic drive. Using Eq.~(\ref{eq:EM_two_state_bound}), we obtain
\begin{eqnarray}
    X_\omega
    &\le&
    \frac{2x_s}{T}
    \int_0^T
    |\cos(\omega t-\varphi)|\,dt=
    \frac{4x_s}{\pi}.
    \label{eq:EM_Xomega_bound}
\end{eqnarray}
Thus, among all periodic responses restricted to the two states
$\pm x_s$, the first harmonic amplitude cannot exceed $4x_s/\pi$.

The bound is attained when the response follows the sign of the fundamental harmonic, corresponding to fully phase-locked switching between the two stable
states
\begin{eqnarray}
    \langle x(t)\rangle_{\rm sq}
    &=&
    x_s\,{\rm sgn}
    [\cos(\omega t-\varphi)]\nonumber\\  &=&
    \frac{4x_s}{\pi}
    \bigg[
        \cos(\omega t-\varphi)
        -
        \frac{1}{3}\cos 3(\omega t-\varphi)
        \nonumber\\
        &&+
        \frac{1}{5}\cos 5(\omega t-\varphi)
        -\cdots
    \bigg].   \label{eq:EM_square_response}
\end{eqnarray}
Hence, the fundamental harmonic has amplitude $4x_s/\pi$, and the
corresponding amplification factor satisfies
\begin{eqnarray}
G_{\rm sq}=\frac{4x_s}{\pi A},    
\end{eqnarray}
 which defines the
two-state square-wave response.
Equality corresponds to fully phase-locked switching, where the system
occupies the drive favored state during each half cycle and switches
once per half period.



\renewcommand{\theequation}{C\arabic{equation}}
\renewcommand{\theHequation}{C.\arabic{equation}}
\setcounter{equation}{0}

\renewcommand{\thefigure}{C\arabic{figure}}
\renewcommand{\theHfigure}{C.\arabic{figure}}
\setcounter{figure}{0}

\begin{figure}[h]
\centering  \includegraphics[width=\linewidth]{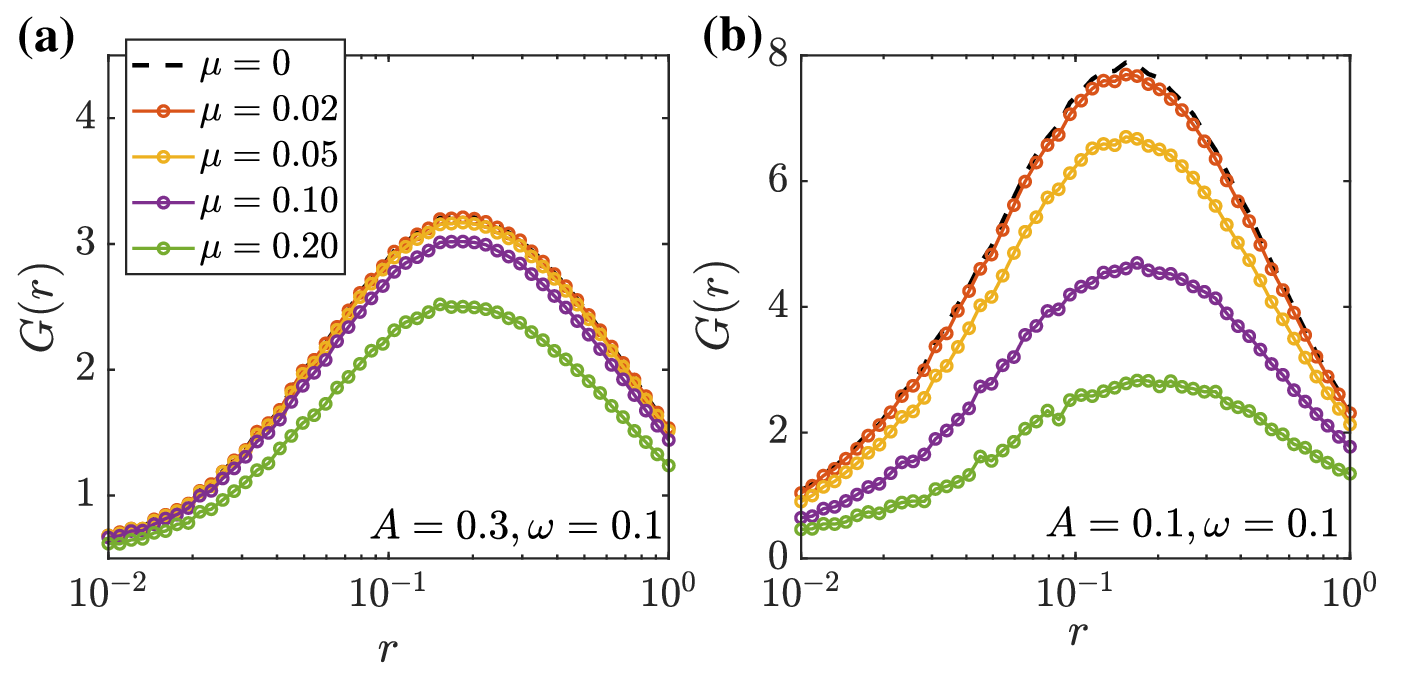}
\caption{
Robustness of SRIR to variations in the reset
position (\ref{eq:robust_reset}). The  amplification factors $G(r)$, obtained by the direct simulations from system (\ref{eq:model}), are shown for different
SR widths $\mu/x_s$ at (a) $A=0.3$ and (b) $A=0.1$, with
$\omega=0.1$. Dashed curves denote ideal SR ($\mu=0$). Parameters: $a=b=x_s=1.$
}
\label{figure6}
\end{figure}

\textit{Robustness to the resetting position.---}
The SRIR does not require resetting precisely to the unstable state of the unforced system. To quantify its
robustness against a finite spread of SR positions around the
unstable state, we replace the ideal SR condition by
\begin{equation}
    x(t_n)=x_r+\mu\eta_n,
    \qquad
    \eta_n\sim{\cal N}(0,1),
    \label{eq:robust_reset}
\end{equation}
where $\mu$ characterizes the spatial uncertainty of the SR
position. As shown in Fig.~\ref{figure6}, the resonant maximum of
$G(r)$ persists for finite $\mu$. Increasing $\mu$ primarily reduces
the peak amplification, whereas the position of the resonance remains
comparatively insensitive. The suppression is more pronounced for
weaker forcing.

This robustness can be understood from the linearized
post-reset dynamics near the unstable state.
For sufficiently slow driving,
Eq.~(\ref{eq:full_post_reset_dynamics}) reduces to
$dx_\phi/d\tau\sim ax_\phi+A\cos\phi$,
with $x_\phi(0)=\mu\eta$, yielding
$
x_\phi(\tau)\sim
\mu\eta e^{a\tau}
+\frac{A\cos\phi}{a}(e^{a\tau}-1).
$
For $a\tau\gg1$, this simplifies to
$x_\phi(\tau)\sim e^{a\tau}
[\mu\eta+(A/a)\cos\phi]$.
Thus, fluctuations in the SR position introduce
a random bias that competes with the phase-dependent
bias of the periodic drive, with their relative
strength characterized by
\begin{equation}
    \epsilon_r=\frac{a\mu}{A}.
    \label{eq:robust_parameter}
\end{equation}
For $\epsilon_r\ll1$, the periodic bias dominates,
and the resonant response is only weakly affected.
As $\epsilon_r$ approaches unity, positional
fluctuations increasingly obscure the phase-dependent
relaxation, explaining the greater sensitivity to
resetting uncertainty at weaker forcing observed
in Fig.~\ref{figure6}.


\begin{thebibliography}{99}

\bibitem{EvansMajumdarSchehr2020} M. R. Evans, S. N. Majumdar, and G. Schehr, Stochastic resetting and applications, J. Phys. A: Math. Theor. \textbf{53}, 193001 (2020).

\bibitem{Pal2022}
A. Pal, S. Kostinski, and S. Reuveni,
The inspection paradox in stochastic resetting,
J. Phys. A: Math. Theor. \textbf{55}, 021001 (2022).

\bibitem{Gupta2022}
S. Gupta and A. M. Jayannavar,
Stochastic resetting: A (very) brief review,
Front. Phys. \textbf{10}, 789097 (2022).

\bibitem{EvansMajumdar2011} M. R. Evans and S. N. Majumdar, Diffusion with stochastic resetting, Phys. Rev. Lett. \textbf{106}, 160601 (2011).

\bibitem{Majumdar2015}
S. N. Majumdar, S. Sabhapandit, and G. Schehr,
Dynamical transition in the temporal relaxation of stochastic
processes under resetting,
Phys. Rev. E \textbf{91}, 052131 (2015).

\bibitem{liang2025} Y. Liang, Q. Wei, W. Wang, and A. G. Cherstvy,
Ultraslow diffusion processes under stochastic resetting,
Phys. Fluids \textbf{37}, 032014 (2025).

\bibitem{sandev2022} T. Sandev, V. Domazetoski, L. Kocarev, R. Metzler, and A. Chechkin,
Heterogeneous diffusion with stochastic resetting,
J. Phys. A: Math. Theor. \textbf{55}, 074003 (2022).


\bibitem{Sokolov2023}
I. M. Sokolov,
Linear response and fluctuation-dissipation relations for Brownian
motion under resetting,
Phys. Rev. Lett. \textbf{130}, 067101 (2023).

\bibitem{sand2022} V. Stojkoski, T. Sandev, L. Kocarev, and A. Pal,
Autocorrelation functions and ergodicity in diffusion with stochastic
resetting, J. Phys. A: Math. Theor. \textbf{55} 104003 (2022).

\bibitem{barkai2023} E. Barkai, R. Flaquer-Galm{\'e}s, and V. M{\'e}ndez,
Ergodic properties of Brownian motion under stochastic resetting,
Phys. Rev. E \textbf{108}, 064102 (2023).

\bibitem{wang2022} W. Wang, A. G. Cherstvy, R. Metzler, and I. M. Sokolov,
Restoring ergodicity of stochastically reset anomalous-diffusion
processes, Phys. Rev. Res. \textbf{4}, 013161 (2022).

\bibitem{Reuveni2016}
S. Reuveni,
Optimal stochastic restart renders fluctuations in first passage
times universal,
Phys. Rev. Lett. \textbf{116}, 170601 (2016).

\bibitem{pal2017} A. Pal and S. Reuveni, First passage under restart,
Phys. Rev. Lett. \textbf{118}, 030603 (2017).

\bibitem{ChechkinSokolov2018}
A. V. Chechkin and I. M. Sokolov,
Random search with resetting: A unified renewal approach,
Phys. Rev. Lett. \textbf{121}, 050601 (2018).

\bibitem{pal2019} A. Pal, I. Eliazar, and S. Reuveni,
First passage under restart with branching,
Phys. Rev. Lett. \textbf{122}, 020602 (2019).

\bibitem{DeBruyne2020}
B. De Bruyne, J. Randon-Furling, and S. Redner,
Optimization in first-passage resetting,
Phys. Rev. Lett. \textbf{125}, 050602 (2020).

\bibitem{reuveni2014} S. Reuveni, M. Urbakh, and J. Klafter, Role
of substrate unbinding in Michaelis-Menten enzymatic reactions,
Proc. Natl. Acad. Sci. USA \textbf{111}, 4391 (2014).


\bibitem{TalFriedman2020} O. Tal-Friedman, A. Pal, A. Sekhon, S. Reuveni, and Y. Roichman, Experimental realization of diffusion with stochastic resetting, J. Phys. Chem. Lett. \textbf{11}, 7350 (2020).

\bibitem{Besga2020}
B. Besga, A. Bovon, A. Petrosyan, S. N. Majumdar, and S. Ciliberto,
Optimal mean first-passage time for a Brownian searcher subjected to
resetting: Experimental and theoretical results,
Phys. Rev. Res. \textbf{2}, 032029(R) (2020).

\bibitem{Faisant2021}
F. Faisant, B. Besga, A. Petrosyan, S. Ciliberto, and S. N. Majumdar,
Optimal mean first-passage time of a Brownian searcher with resetting
in one and two dimensions: Experiments, theory and numerical tests,
J. Stat. Mech. \textbf{11},  113203 (2021).


\bibitem{Vatash2025}
R. Vatash and Y. Roichman,
Many-body colloidal dynamics under stochastic resetting:
Competing effects of particle interactions on the steady-state distribution,
Phys. Rev. Res. \textbf{7}, L032020 (2025).

\bibitem{Goerlich2024}
R. Goerlich, T. D. Keidar, and Y. Roichman,
Resetting as a swift equilibration protocol in an anharmonic potential,
Phys. Rev. Res. \textbf{6}, 033162 (2024).

\bibitem{Olsen2026}
K. S. Olsen, H. L\"owen, and L. Caprini,
Optimal area exploration by resetting active particles,
Phys. Rev. Lett. \textbf{137}, 057102 (2026).

\bibitem{Olsen2024}
K. S. Olsen, D. Gupta, F. Mori, and S. Krishnamurthy,
Thermodynamic cost of finite-time stochastic resetting,
Phys. Rev. Res. \textbf{6}, 033343 (2024).

\bibitem{Gupta2020}
D. Gupta, C. A. Plata, and A. Pal,
Work fluctuations and Jarzynski equality in stochastic resetting,
Phys. Rev. Lett. \textbf{124}, 110608 (2020).


\bibitem{meyl2015} J. M. Meylahn, S. Sabhapandit, and H. Touchette,
Large deviations for Markov processes with resetting, Phys. Rev. E
\textbf{92}, 062148 (2015).

\bibitem{boyer2021} F. H. Gonzalez, A. P. Riascos, and D. Boyer,
Diffusive transport on networks with stochastic resetting to multiple
nodes, Phys. Rev. E \textbf{103}, 062126 (2021).

\bibitem{McNamara1989} B. McNamara and K. Wiesenfeld, Theory of stochastic resonance, Phys. Rev. A \textbf{39}, 4854 (1989).

\bibitem{Gammaitoni1998} L. Gammaitoni, P. H\"anggi, P. Jung, and F. Marchesoni, Stochastic resonance, Rev. Mod. Phys. \textbf{70}, 223 (1998).

\bibitem{Wellens2004}
T. Wellens, V. Shatokhin, and A. Buchleitner,
Stochastic resonance,
Rep. Prog. Phys. \textbf{67}, 45 (2004).

\bibitem{Pikovsky2001}
A. Pikovsky, M. Rosenblum, and J. Kurths,
\textit{Synchronization: A Universal Concept in Nonlinear Sciences}
(Cambridge University Press, Cambridge, 2001).

\bibitem{Benzi1981}
R. Benzi, A. Sutera, and A. Vulpiani,
The mechanism of stochastic resonance,
J. Phys. A: Math. Gen. \textbf{14}, L453 (1981).

\bibitem{Benzi1982}
R. Benzi, G. Parisi, A. Sutera, and A. Vulpiani,
Stochastic resonance in climatic change,
Tellus \textbf{34}, 10 (1982).




\bibitem{FauveHeslot1983}
S. Fauve and F. Heslot,
Stochastic resonance in a bistable system,
Phys. Lett. A \textbf{97}, 5 (1983).

\bibitem{McNamaraRoy1988}
B. McNamara, K. Wiesenfeld, and R. Roy,
Observation of stochastic resonance in a ring laser,
Phys. Rev. Lett. \textbf{60}, 2626 (1988).

\bibitem{MantegnaSpagnolo1994}
R. N. Mantegna and B. Spagnolo,
Stochastic resonance in a tunnel diode,
Phys. Rev. E \textbf{49}, R1792 (1994).

\bibitem{LevinMiller1996}
J. E. Levin and J. P. Miller,
Broadband neural encoding in the cricket cercal sensory system
enhanced by stochastic resonance,
Nature \textbf{380}, 165 (1996).

\bibitem{Douglass1993}
J. K. Douglass, L. Wilkens, E. Pantazelou, and F. Moss,
Noise enhancement of information transfer in crayfish
mechanoreceptors by stochastic resonance,
Nature \textbf{365}, 337 (1993).

\bibitem{Pikovsky1997}
A. Pikovsky and J. Kurths,
Coherence resonance in a noise-driven excitable system,
Phys. Rev. Lett. \textbf{78}, 775 (1997).


\bibitem{Gang1993}
G. Hu, T. Ditzinger, C. Z. Ning, and H. Haken,
Stochastic resonance without external periodic force,
Phys. Rev. Lett. \textbf{71}, 807 (1993).

\bibitem{Longtin1997}
A. Longtin,
Autonomous stochastic resonance in bursting neurons,
Phys. Rev. E \textbf{55}, 868 (1997).

\bibitem{Lenton2008}
T. M. Lenton, H. Held, E. Kriegler,
J. W. Hall, W. Lucht, S. Rahmstorf,
and H. J. Schellnhuber,
Tipping elements in the Earth's climate system,
Proc. Natl. Acad. Sci. USA \textbf{105},
1786 (2008).

\bibitem{Brette2005}
R. Brette and W. Gerstner,
Adaptive exponential integrate-and-fire model as
an effective description of neuronal activity,
J. Neurophysiol. \textbf{94}, 3637 (2005).

\bibitem{Vinod2022}
D. Vinod, A. G. Cherstvy, W. Wang, R. Metzler, and I. M. Sokolov,
Nonergodicity of reset geometric Brownian motion,
Phys. Rev. E \textbf{105}, L012106 (2022).

\bibitem{Stojkoski2021}
V. Stojkoski, T. Sandev, L. Kocarev, and A. Pal,
Geometric Brownian motion under stochastic resetting:
A stationary yet nonergodic process,
Phys. Rev. E \textbf{104}, 014121 (2021).

\bibitem{footnote1} The threshold follows from the saddle-node conditions
$ax_c-bx_c^3+F_c=0$ and $a-3bx_c^2=0$, yielding
$A_c=|F_c|=2a^{3/2}/(3\sqrt{3b})$.

\bibitem{footnote2} In the long-time periodic state, the initial drive phase $\phi_0$
only multiplies the $n$th harmonic coefficient by the phase factor
$e^{in\phi_0}$ and therefore does not affect its amplitude
$|C_n(r)|$. Hence, without a loss of generality, we set $\phi_0=0$.

\bibitem{SM}
See Supplemental Material for the general renewal
formulation, coarse-grained reduction, optimal-rate
scaling beyond bistability, selective control of
SRIR and NIR, numerical details, and Movies S1 and S2
illustrating the reset-induced dynamics in bistable
and tristable systems, respectively.

\bibitem{zhu2024}
J. Zhu, F. Zhao, Y. Li, and X. Liu,
Rotational stochastic resonance in multistable systems,
Physica A \textbf{634}, 129473 (2024).

\bibitem{Risken1989}
H. Risken,
\textit{The Fokker--Planck Equation: Methods of Solution and Applications},
2nd ed., Springer Series in Synergetics Vol.~18
(Springer-Verlag, Berlin, 1989).

\bibitem{Kramers1940}
H. A. Kramers,
Brownian motion in a field of force and the diffusion model of chemical reactions,
Physica \textbf{7}, 284 (1940).

\bibitem{Hanggi1990}
P. H\"anggi, P. Talkner, and M. Borkovec,
Reaction-rate theory: Fifty years after Kramers,
Rev. Mod. Phys. \textbf{62}, 251 (1990).

\bibitem{Gammaitoni1989} L. Gammaitoni, F. Marchesoni, E. Menichella-Saetta, and S. Santucci, Stochastic resonance in bistable systems, Phys. Rev. Lett. \textbf{62}, 349 (1989).

\bibitem{Jop2008} P. Jop, A. Petrosyan, and S. Ciliberto, Work and dissipation fluctuations near the stochastic resonance of a colloidal particle, Europhys. Lett. \textbf{81}, 50005 (2008).

\bibitem{Bhowmik2014}
D. Bhowmik, L. You, and S. Salahuddin,
Spin Hall effect clocking of nanomagnetic logic without a magnetic field,
Nat. Nanotechnol. \textbf{9}, 59 (2014).


\bibitem{Jolakoski2025}
P. Jolakoski, L. Basnarkov, L. Kocarev,
A. Popovska-Mitrovikj, V. Bakeva, and T. Sandev,
Response to an external field of a generalized Langevin equation
with stochastic resetting of the memory kernel,
Phys. Rev. E \textbf{112}, 024120 (2025).

\bibitem{DiBello2023}
C. Di Bello, A. V. Chechkin, A. K. Hartmann,
Z. Palmowski, and R. Metzler,
Time-dependent probability density function for partial resetting dynamics,
New J. Phys. \textbf{25}, 082002 (2023).

\bibitem{Bodrova2019}
A. S. Bodrova, A. V. Chechkin, and I. M. Sokolov,
Scaled Brownian motion with renewal resetting,
Phys. Rev. E \textbf{100}, 012120 (2019).

\bibitem{Gupta2019}
D. Gupta,
Stochastic resetting in underdamped Brownian motion,
J. Stat. Mech. \textbf{3}  033212 (2019).

\bibitem{Franke2026}
H. Franke and I. M. Sokolov,
Resetting of underdamped Brownian motion,
Phys. Rev. E \textbf{113}, 064125 (2026).

\bibitem{code}
See the numerical codes available at Zenodo,
\url{https://doi.org/10.5281/zenodo.23798786}.

\end{thebibliography}
\end{document}


\setcounter{secnumdepth}{2}
\makeatletter
\renewcommand{\p@subsection}{}
\makeatother
\title{Supplemental Material for "Weak-Signal Amplification through Stochastic Resetting"}

\maketitle

This Supplemental Material provides the theoretical and numerical
details supporting the main text. Sec.~\ref{sm-sec-1} develops a general
renewal formulation for periodically driven systems under stochastic resetting (SR). Sec.~\ref{sm-sec-2} introduces a
phase-dependent coarse-grained approximation, identifies the competing
response losses underlying the optimal SR rate, and derives the
low-frequency scaling of the optimal SR rate and discusses the generality beyond bistable system. 
Sec.~\ref{sm-sec-3} discusses how SR-induced resonance (SRIR) and noise-induced resonance (NIR) can be selectively controlled. Sec.~\ref{sm-sec-4} describes the numerical methods used for the
SR dynamics, renewal calculations, and the comparison of SRIR with NIR. Finally, Sec.~\ref{sec:movies} describes Supplemental Movies S1 and S2, which illustrate the SRIR dynamics in bistable and tristable systems, respectively.

\begingroup
\small
\tableofcontents
\endgroup

\section{General renewal formulation}
\label{sm-sec-1}

\subsection{General periodically driven dynamics and renewal response}
\label{sm:gen:renewal_section}

The exact response can be obtained without specifying a potential.
Consider a $d$-dimensional periodic dynamical system
\begin{subequations}\label{sm:gen:dynamics}
\begin{eqnarray}
\frac{d\mathbf x}{dt}&=&\boldsymbol{\mathcal F}(\mathbf x,\theta;A),\\
\frac{d\theta}{dt} &=& \omega,\\
\boldsymbol{\mathcal F}
(\mathbf x,\theta;A)
&=&
\boldsymbol{\mathcal F}(\mathbf x,\theta+2\pi;A),
\end{eqnarray}   
\end{subequations}
where $\theta$ is the driving phase, $\omega$ is the angular
frequency, and $A$ parametrizes the amplitude of periodic forcing.
At Poisson times with rate $r>0$, the full dynamical state is redrawn
from a normalized distribution $\rho_R(\mathbf x)$, independently of
its pre-reset value. The external phase $\theta$ is not reset. Neither the
inter-reset dynamics nor $\rho_R$ depends on $r$. Fixed-point SR
corresponds to $\rho_R(\mathbf x)=\delta(\mathbf x-\mathbf x_r)$.

For a reset occurring at time $t_r$, we define the reset phase
$\phi=\phi_0+\omega t_r\ ({\rm mod}\ 2\pi)$ and the elapsed time
$\tau=t-t_r$ since that reset. We denote by
$\mathbf{X}_\phi(\tau;\mathbf{x}_0)$ the uninterrupted trajectory at
elapsed time $\tau$, initialized at $\mathbf{x}_0$ immediately after
a reset occurring at phase $\phi$. It satisfies
\begin{subequations}\label{sm:gen:trajectory}
\begin{align}
&\frac{d\mathbf X_\phi}{d\tau}
=\boldsymbol{\mathcal F}
 (\mathbf X_\phi,\phi+\omega\tau;A),\\
&\mathbf X_\phi(0;\mathbf x_0)=\mathbf x_0.
\end{align}   
\end{subequations}
For a real observable $Q$, we define the uninterrupted
post-reset response as
\begin{equation}
q_\phi(\tau)=\int d\mathbf x_0\,\rho_R(\mathbf x_0)
 Q[\mathbf X_\phi(\tau;\mathbf x_0)].
\label{sm:gen:postreset_observable}
\end{equation}
Here assume that $Q$ is bounded on the domain of reachable states.
This is sufficient for convergence of the integrals and limiting operations below.
All properties of the uninterrupted dynamics enter through
$q_\phi(\tau)$.

Similarly, for an initial state distribution
$\rho_{\rm init}(\mathbf{x}_0)$, the uninterrupted response from the
initial preparation is
\begin{equation}
    q_{\rm init}(t)
    =
    \int d\mathbf{x}_0\,
    \rho_{\rm init}(\mathbf{x}_0)
    Q[\mathbf{X}_{\phi_0}(t;\mathbf{x}_0)].
    \label{eq:q_init}
\end{equation}
When the initial and reset-state distributions coincide,
$\rho_{\rm init}=\rho_R$, one has
$q_{\rm init}(t)=q_{\phi_0}(t)$.

At time $t$, the probability that no reset has occurred is
$e^{-rt}$. For $0<\tau<t$, the probability density that the most
recent reset occurred at time $(t-\tau)$ is $re^{-r\tau}$.
Let
$\mathcal{Q}(t)=\langle Q[\mathbf{x}(t)]\rangle$
denote the full ensemble-averaged response. Conditioning this on the most
recent reset gives
\begin{equation}
    \mathcal{Q}(t)
    =
    e^{-rt}q_{\rm init}(t)
    +
    r\int_0^t e^{-r\tau}
    q_{\phi_0+\omega(t-\tau)}(\tau)\,d\tau .\label{sm:gen:finite_renewal}
\end{equation}
The reset-phase argument is always understood modulo $2\pi$.
In the long-time limit, the response approaches the periodic form
\begin{equation}
\mathcal Q_{\rm per}(t)
=
r\int_0^\infty
e^{-r\tau}
q_{\phi_0+\omega(t-\tau)}(\tau)\,d\tau .
\label{sm:gen:periodic_renewal}
\end{equation}
For any bounded observable $|Q|\leq Q_{\max}$, the finite-time response (\ref{sm:gen:finite_renewal})
converges exponentially to this periodic state,
\begin{equation}
\left|
\mathcal Q(t)-\mathcal Q_{\rm per}(t)
\right|
\leq
2Q_{\max}e^{-rt}.
\end{equation}
Thus, the periodic
response follows directly from the renewal construction without
requiring the uninterrupted system to possess a unique attractor.

Since $q_\phi(\tau)$ is $2\pi$-periodic in the reset phase $\phi$, we can use the Fourier series
\begin{equation}
q_\phi(\tau)
=
\sum_{n=-\infty}^{\infty}
h_n(\tau)e^{in\phi},
\label{eq:q_fourier}
\end{equation}
where
\begin{equation}
h_n(\tau)
=
\frac{1}{2\pi}
\int_0^{2\pi}
q_\phi(\tau)e^{-in\phi}\,d\phi .
\label{sm:gen:hn}
\end{equation}
Substituting Eq.~(\ref{eq:q_fourier}) into
Eq.~(\ref{sm:gen:periodic_renewal}) gives
\begin{equation}
\mathcal Q_{\rm per}(t)
=
\sum_{n=-\infty}^{\infty}
C_n(r;\phi_0)e^{in\omega t},
\label{eq:Q_fourier}
\end{equation}
where
\begin{align}
C_n(r;\phi_0)
&=
e^{in\phi_0}
r\int_0^\infty
e^{-(r+in\omega)\tau}
h_n(\tau)\,d\tau
\nonumber\\
&=
e^{in\phi_0}
r\widetilde h_n(r+in\omega),
\label{eq:full_Cn_laplace}
\end{align}
with
\begin{equation}
\widetilde h_n(s)
=
\int_0^\infty
e^{-s\tau}h_n(\tau)\,d\tau .
\label{eq:full_hn_laplace}
\end{equation}
We note that the initial drive phase $\phi_0$ only introduces the
phase factor $e^{in\phi_0}$ in the $n$th harmonic and therefore does
not affect its amplitude. Without loss of generality, we set
$\phi_0=0$ hereafter \begin{eqnarray}\label{eq-s12}
C_n(r;\phi_0=0):=C_n(r).
\end{eqnarray}

Equations~(\ref{sm:gen:hn})--(\ref{eq:full_hn_laplace}) provide the general renewal representation of the coherent
harmonics. Its structure is independent of the detailed deterministic
dynamics, which enters entirely through $h_n(\tau)$. This is used in Eq.~(4) of the main text.

\subsection{Amplification factor}
\label{sm:int:specialization}

For the fundamental harmonic, the general renewal relation with the help of Eqs~(\ref{eq-s12}) and (\ref{eq:full_hn_laplace}) gives
\begin{equation}
C_1(r)
=
r\widetilde h_1(r+i\omega).
\label{eq:general_C1}
\end{equation}
The coherent first harmonic amplitude of the observable $Q$ is
\begin{equation}
Q_\omega(r)=2|C_1(r)|.
\label{eq:general_output_amplitude}
\end{equation}
We define the corresponding amplification factor $G$ as the ratio of
the output amplitude $Q_\omega(r)$ to the forcing amplitude $A$,
\begin{equation}
G(r)
\equiv
\frac{Q_\omega(r)}{A}
=
\frac{2|C_1(r)|}{A}
=
\frac{2r}{A}
\left|
\widetilde h_1(r+i\omega)
\right|.
\label{eq:general_gain}
\end{equation}
Using the definition of $h_1(\tau)$, the amplification factor can equivalently be written
as
\begin{equation}
G(r)
=
\frac{r}{\pi A}
\left|
\int_0^{2\pi}d\phi
\int_0^\infty d\tau\,
q_\phi(\tau)
e^{-r\tau-i(\phi+\omega\tau)}
\right|.
\label{eq:general_gain_double_integral}
\end{equation}
The detailed uninterrupted dynamics enters entirely through
$q_\phi(\tau)$, while the resetting statistics enter through the
exponential age weight $re^{-r\tau}$. These results and definitions are used for Eq.~(5) in the main text.

\subsection{Exact optimal condition}
\label{sm:int:exact_optimum}

For either a general observable or the bistable position response, Eq.~(\ref{eq:general_gain}) is valid. Taking the logarithmic
derivative with respect to $r$ gives
\begin{eqnarray}
\frac{d\ln G}{dr}
&=&
\frac{1}{r}
+
\operatorname{Re}
\left[
\frac{\partial_r\widetilde h_1(r+i\omega)}
{\widetilde h_1(r+i\omega)}
\right],
\label{eq:full_matching_derivative}
\end{eqnarray} where we used
$d\ln|Z|/dr=\operatorname{Re}(Z'/Z)$ for a nonzero complex
function $Z(r)$.
Since $h_1(\tau)$ is independent of $r$, differentiation under the
integral yields
\begin{eqnarray}
\partial_r\widetilde h_1(r+i\omega)
=
-\int_0^\infty
\tau e^{-(r+i\omega)\tau}h_1(\tau)\,d\tau .
\label{eq:full_h1_derivative}
\end{eqnarray}
Therefore,
\begin{equation}
\frac{d\ln G}{dr}
=
\frac{1}{r}
-
\tau_{\rm eff}(r,\omega),
\label{eq:full_matching_pre}
\end{equation}
where
\begin{equation}
\tau_{\rm eff}(r,\omega)
=
\operatorname{Re}
\frac{\displaystyle\int_0^\infty
\tau e^{-(r+i\omega)\tau}h_1(\tau)\,d\tau}
{\displaystyle\int_0^\infty
e^{-(r+i\omega)\tau}h_1(\tau)\,d\tau}.
\label{eq:full_tau_eff}
\end{equation} Every interior stationary point for variable $r$ satisfies
\begin{equation}
r_*\tau_{\rm eff}(r_*,\omega)=1.
\label{eq:full_matching}
\end{equation}
For the bistable model, the optimal rate is selected by
\begin{equation}
r_{\rm opt}\in\operatorname*{arg\,max}_{r\geq0}G(r)
\label{eq:full_ropt_definition}
\end{equation}
using the appropriate no-reset response at the boundary $r=0$.

If the maximum is interior, it must obey
Eq.~(\ref{eq:full_matching}); roots of that condition must be checked
against one another and against boundary values. Numerical integration
of the exact response, followed by one-dimensional optimization or
root finding with this check, yields the prediction without an
additional dynamical approximation.  Equation~(\ref{eq:full_matching}) is only a condition for an interior
extremum; it neither guarantees that such an extremum exists nor that
it is a maximum. The following simple example illustrates this point.

The existence of an interior optimum is not guaranteed by the renewal
structure alone. As a simple counterexample, consider the driven linear equation
\begin{equation}
\frac{dx(t)}{dt}=-\kappa x(t)+A\cos(\omega t),
\label{sm:linear_relaxator} 
\end{equation} where $\kappa>0$ and the reset position is $x_r=0$.
For a reset occurring at phase $\phi$, the uninterrupted trajectory is
\begin{equation}
x_\phi(\tau)
=
A\,{\rm Re}\left[
e^{i\phi}
\frac{e^{i\omega\tau}-e^{-\kappa\tau}}
{\kappa+i\omega}
\right],
\end{equation}
so that
\begin{equation}
h_1(\tau)
=
\frac{A}{2}
\frac{e^{i\omega\tau}-e^{-\kappa\tau}}
{\kappa+i\omega}.
\end{equation}
Substitution this into the exact renewal relation
Eqs.~(\ref{eq:general_C1}) and (\ref{eq:general_gain}) gives
\begin{equation}
C_1(r)
=
\frac{A}{2(r+\kappa+i\omega)},
\qquad
G(r)
=
\frac{1}{\sqrt{(r+\kappa)^2+\omega^2}}.
\label{sm:linear_gain}
\end{equation}
Hence $dG/dr<0$ for all $r\geq0$, and the maximum occurs at the
boundary $r_{\rm opt}=0$. Thus, a finite post-reset relaxation time
alone is insufficient to produce SRIR.
An interior optimum requires additional response structure, such as
the competition between reset-triggered phase selection and finite
post-reset relaxation, as developed in Sec.~\ref{sm-sec-2}.

\section{Coarse-grained reduction}
\label{sm-sec-2}

The exact renewal formulation developed in
Sec.~\ref{sm-sec-1} applies to general periodically
driven systems under Poisson resetting, without
presupposing or guaranteeing resonance.
We now identify a more restrictive response-level
structure that gives rise to competing effects
responsible for an optimal resetting rate.
This construction requires neither bistability
nor multistability; it does not assume resetting
near an unstable state either.
Local dynamical instability provides one natural
mechanism for realizing such a response structure.

\subsection{Optimal rate scaling}
\label{sm:int:loss_optimum}

Suppose that a reset occurring at forcing phase $\phi$ selects a
well-defined post-reset response $Q_{\rm sel}(\phi)$ where the subscript "sel" denotes "selection". This response is established after a finite, phase-dependent relaxation time $\tau_s(\phi)$. At the level of the observable entering the coherent
response, the reduction is
\begin{equation}
q_\phi(\tau)
=
Q_{\rm sel}(\phi)
\Theta[\tau-\tau_s(\phi)].
\label{eq:general_step_response}
\end{equation}
Here $\Theta[z]$ is the Heaviside function, $Q_{\rm sel}(\phi)$ denotes the selected observable output and
$\tau_s(\phi)$ characterizes the time required for this output to
develop following a reset. We note that Eq.~(\ref{eq:general_step_response}) is a coarse-grained approximation in which the transient evolution toward the selected state is neglected, and the observable is assumed to reach its selected value only after the relaxation time $\tau_s(\phi)$.

Equation~(\ref{eq:general_step_response}) is a response-level
description and does not require a particular state-space geometry.
For a system with $K$ phase-selectable branches,
$Q_{\rm sel}(\phi)$ may take the values
$\{Q_1,\ldots,Q_K\}$. The essential requirement is that, once established,
the selected output remains approximately persistent on time scale
relevant to the renewal response. Systems whose post-reset observable
continues to undergo an intrinsic oscillation after the initial
relaxation need not belong to this class, even though the exact renewal
formulation in Sec.~S1 remains applicable.

Substituting Eq.~(\ref{eq:general_step_response}) into the general
renewal representation (\ref{eq:general_gain}) gives
\begin{subequations}
\begin{align}
G(r)
&=
\frac{2r}{A\sqrt{r^2+\omega^2}}
|\mathcal M(r+i\omega)|,
\label{sm:gen:step_harmonic}\\
\mathcal M(z)
&=
\frac{1}{2\pi}
\int_0^{2\pi}
Q_{\rm sel}(\phi)e^{-i\phi}
e^{-z\tau_s(\phi)}\,d\phi.
\label{sm:gen:M}
\end{align}    
\end{subequations}
These relations are exact within the reduced step-response
description. The factor $r/(r+i\omega)$ describes the renewal of the
reset phase, whereas $\mathcal M(z)$ contains the phase-selected
post-reset relaxation dynamics.

To characterize $\mathcal M(z)$, we introduce the response weighted moments
\begin{equation}
M_k
=
\frac{1}{2\pi}
\int_0^{2\pi}
Q_{\rm sel}(\phi)e^{-i\phi}
\tau_s^k(\phi)\,d\phi,
\quad k=\{0,1,\ldots\} ,
\label{sm:gen:moments}
\end{equation}
which are generally complex. We assume $M_0\neq0$ and write
\begin{equation}
\frac{M_1}{M_0}
=
\tau_b+i\tau_p,
\label{sm:gen:complex_delay}
\end{equation}
where
\begin{equation}
\tau_b
\equiv
{\rm Re}\!\left(\frac{M_1}{M_0}\right),\qquad  \tau_p=\operatorname{Im}\left(\frac{M_1}{M_0}\right).
\label{sm:gen:tau_b}
\end{equation}
We refer to $\tau_b$ as the \textit{effective relaxation time}.
It differs from an ordinary phase-averaged relaxation time because the
weight $Q_{\rm sel}(\phi)e^{-i\phi}$ retains specifically those
post-reset events that contribute coherently to the fundamental
harmonic. A positive $\tau_b$ means that finite post-reset relaxation
produces a leading suppression of the coherent response as the
resetting rate increases, as shown below.

For fixed forcing parameters, both $Q_{\rm sel}(\phi)$ and
$\tau_s(\phi)$ are determined entirely by the uninterrupted post-reset
dynamics and are therefore independent of the resetting rate $r$.
Consequently, $\tau_b$ is also independent of $r$, although it may in
general depend on $A$, $\omega$, and other system
parameters.

To control higher-order corrections, we introduce the second absolute
weighted moment
\begin{equation}
\widehat M_2
=
\frac{1}{2\pi}
\int_0^{2\pi}
|Q_{\rm sel}(\phi)|\tau_s^2(\phi)\,d\phi
\label{sm:gen:second_moment}
\end{equation}
and define
\begin{equation}
\chi_2
=
\frac{\widehat M_2}{|M_0|}
+
\left|
\frac{M_1}{M_0}
\right|^2.
\label{sm:gen:chi2}
\end{equation}
We focus on phase-selected responses satisfying the convergence
\begin{equation}
\widehat M_2<\infty.
\label{sm:gen:scaling_conditions}
\end{equation}
The first condition ensures a positive relaxation loss with increasing
resetting rate, while the second controls the higher-order corrections.
In particular, for $\widehat M_2<\infty$, the transform $\mathcal M(z)$
admits, as $z\to0$,
\begin{equation}
\mathcal M(z)
=
M_0-zM_1
+
O\!\left(|z|^2\widehat M_2\right).
\end{equation}

Setting $z=r+i\omega$ gives
\begin{equation}
\frac{|\mathcal M(r+i\omega)|}{|M_0|}
=
1-r\tau_b+\omega\tau_p
+
O\!\left[
(r^2+\omega^2)\chi_2
\right].
\label{sm:gen:M_modulus}
\end{equation}
The term $\omega\tau_p$ modifies the low-frequency reference response,
but it is independent of $r$ to this order. In contrast, $r\tau_b$ is the
leading relaxation loss relevant to optimization over the resetting
rate.

From Eq.~(\ref{sm:gen:step_harmonic}), it follows that
\begin{equation}
\frac{G(r)}{G_0}
=
\frac{r}{\sqrt{r^2+\omega^2}}
\frac{|\mathcal M(r+i\omega)|}{|M_0|},
\qquad
G_0=\frac{2|M_0|}{A}.
\label{sm:gen:normalized_gain}
\end{equation}
For any finite $\omega>0$, sufficiently rare resetting gives
\begin{equation}
\frac{r}{\sqrt{r^2+\omega^2}}
\sim
\frac{r}{\omega}
\to0,
\label{sm:gen:rare_reset}
\end{equation}
so that the resetting-generated coherent response is suppressed because
phase-selection events occur too infrequently.

At larger resetting rates, finite post-reset relaxation produces the
competing loss $r\tau_b$. For slow driving, these two limits leave an
intermediate asymptotic window
\begin{equation}
\omega\ll r\ll\tau_b^{-1},
\label{sm:gen:intermediate_regime}
\end{equation}
where $\tau_b$ is the effective relaxation time defined in Eq.~(\ref{sm:gen:tau_b}). Within this regime, we have
\begin{equation}
\frac{r}{\sqrt{r^2+\omega^2}}
=
1-\frac{\omega^2}{2r^2}
+
O\!\left(
\frac{\omega^4}{r^4}
\right).
\end{equation}
Combining this expression with Eq.~(\ref{sm:gen:M_modulus}) gives
\begin{align}
\frac{G(r)}{G_0}
={}&
1+\omega\tau_p
-\frac{\omega^2}{2r^2}
-r\tau_b
\nonumber\\
&+
O\!\left[
\frac{\omega^4}{r^4}
+
(r^2+\omega^2)\chi_2
+
\frac{\omega^2}{r^2}
\left(
r\tau_b+|\omega\tau_p|
\right)
\right].
\label{sm:gen:remainder}
\end{align}
The two leading $r$-dependent losses in Eq.~(\ref{sm:gen:remainder}) are therefore
\begin{equation}
L(r)
=
\frac{\omega^2}{2r^2}
+
r\tau_b.
\label{sm:int:loss_sum}
\end{equation}
The first is the finite-rate phase-selection loss: resetting events
that are too rare cannot update the selected response sufficiently
rapidly relative to the external phase. The second is the interruption
loss: overly frequent resets suppress the finite post-reset response
relaxation. This is one of the main technical sides of the paper.

As
\begin{equation}
L'(r)
=
-\frac{\omega^2}{r^3}
+\tau_b,
\qquad
L''(r)
=
\frac{3\omega^2}{r^4}>0,
\end{equation}
the leading loss has a unique minimum at
\begin{equation}
r_{\rm opt}(\omega)
\sim
\left(
\frac{\omega^2}{\tau_b}
\right)^{1/3}.
\label{sm:gen:optimal_scaling}
\end{equation}
If the higher-order
moments remaining suitably bounded, Eq.~(\ref{sm:gen:scaling_conditions}),
one can yield the asymptotic scaling
$r_{\rm opt}\propto\omega^{2/3}$.
The exponent therefore follows from the competing
response losses rather than from a specific
potential or resetting location. This is one of the key results, claimed in Eq.~(11) of the main text.

\subsection{Bistable realization}
\label{sm:int:bistable_specialization}
We now apply the general coarse-grained reduction to the bistable
system Eqs.~(1) and (2) of in the main text. The reduced post-reset response is
\begin{equation}
x_\phi(\tau)
=
\sigma_\phi x_s
\Theta[\tau-\tau_s(\phi)],
\label{sm:bistable_step}
\end{equation}
where $\sigma_\phi=\pm1$ denotes the selected well and
$\tau_s(\phi)$ is the corresponding reset-to-well relaxation time.

For weak forcing and sufficiently slow driving, near the unforced unstable state, the post-reset dynamics becomes 
\begin{equation}
\frac{dx_\phi(\tau)}{d\tau}
\sim ax_\phi(\tau)+A\cos\phi
\label{eq:rho1_qs_force}
\end{equation}
with the solution
\begin{equation}
x_\phi(\tau)
\sim
\frac{A\cos\phi}{a}
\left(e^{a\tau}-1\right).
\label{sm:int:quartic_local}
\end{equation}
This gives the approximate selection rule of the response (\ref{sm:bistable_step})
\begin{equation}
\sigma_\phi
\sim
\operatorname{sgn}(\cos\phi).
\label{eq:rho1_sigma}
\end{equation}

Using $Q_{\rm sel}(\phi)=x_s\operatorname{sgn}(\cos\phi)$ in the
general moments (\ref{sm:gen:moments}) to (\ref{sm:gen:tau_b}), yields
\begin{equation}
\tau_b
=
\overline{\tau}_s
=
\frac14
\int_0^{2\pi}
|\cos\phi|\tau_s(\phi)\,d\phi .
\label{sm:int:time_mapping}
\end{equation}
Thus, the effective relaxation time $\tau_b$ reduces to the first-harmonic-weighted reset-to-well
time $\overline{\tau}_s$.

For motion toward the selected well, let
\begin{equation}
F_\phi=A|\cos\phi|,
\qquad
\epsilon_\phi=\frac{F_\phi}{ax_s}.
\label{eq:rho1_Fphi}
\end{equation}
The relaxation time of the system (2) in the main text is
\begin{eqnarray}
\tau_s(\phi)
&\sim&
\int_0^{x_s}
\frac{dx}{ax-bx^3+F_\phi}
\nonumber\\
&=&
\frac1a
\int_0^1
\frac{dy}{y-y^3+\epsilon_\phi},
\label{eq:rho1_tau_scaled}
\end{eqnarray}
where $y=x/x_s$. For $\epsilon_\phi\ll1$, the neighborhoods of both
$y=0$ and $y=1$ contribute to the reset-to-well time $\tau_s(\phi)$ logarithmically, yielding
\begin{equation}
\tau_s(\phi)
\sim
\frac{3}{2a}
\ln\left(
\frac{ax_s}{A|\cos\phi|}
\right).
\label{eq:rho1_tau_asymptotic}
\end{equation}
The factor $3/(2a)$ contains both the unstable departure contribution
$a^{-1}\ln(1/\epsilon_\phi)$ and the near-well contribution
$(2a)^{-1}\ln(1/\epsilon_\phi)$.

The logarithmic divergence at the isolated phase-selection boundaries
$\cos\phi=0$ is integrable, so the second absolute weighted moment
used in Sec.~S2.1 remains finite. Substituting
Eq.~(\ref{eq:rho1_tau_asymptotic}) into
Eq.~(\ref{sm:int:time_mapping}) gives
\begin{equation}
\overline{\tau}_s
\sim
\frac{3}{2a}
\left[
\ln\left(
\frac{ax_s}{2A}
\right)
+
1
\right].
\label{eq:rho1_mu1}
\end{equation}
Hence $\tau_b=\overline{\tau}_s>0$, and the condition (\ref{sm:gen:scaling_conditions}) required by
the general phase-selected reduction, is satisfied.

Applying the amplification factor  (\ref{sm:gen:remainder})
gives
\begin{equation}
\frac{G(r)}{G_{0}}
\sim
1
-\frac{\omega^2}{2r^2}
-r\overline{\tau}_s,\quad  G_0
=
\frac{4x_s}{\pi A},
\label{eq:three_state_gain_expansion}
\end{equation}
and therefore taking the maximum of $G(r)$ yields
\begin{equation}
r_{\rm opt}(\omega)
\sim
\left(
\frac{\omega^2}{\overline{\tau}_s}
\right)^{1/3}.
\label{eq:rho1_ropt_asymptotic}
\end{equation}
The scaling $r_{\rm opt}(\omega)
\propto
\omega^{2/3}$ follows from the general two-loss balance, whereas
the logarithmic amplitude dependence $\overline{\tau}_s(A)$ originates from the specific
reset-to-well relaxation of the bistable dynamics. Here, we refer to the description of Fig.~3 in the main text.

\subsection{Three-stable-state realization}
\label{sm:three_attractor}

\begin{figure}
\centering
(a)\includegraphics[width=0.9\linewidth]{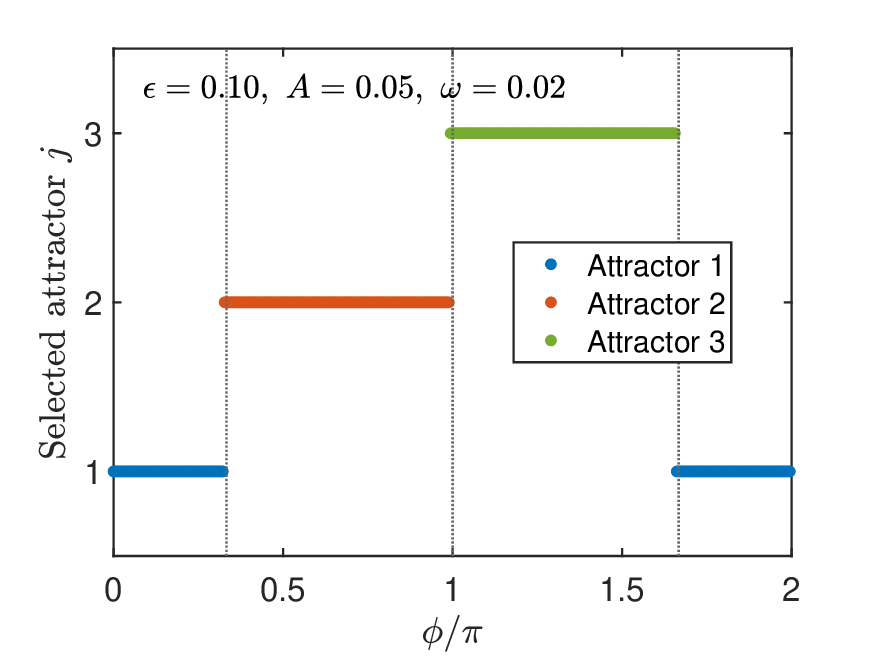}
(b)\includegraphics[width=0.9\linewidth]{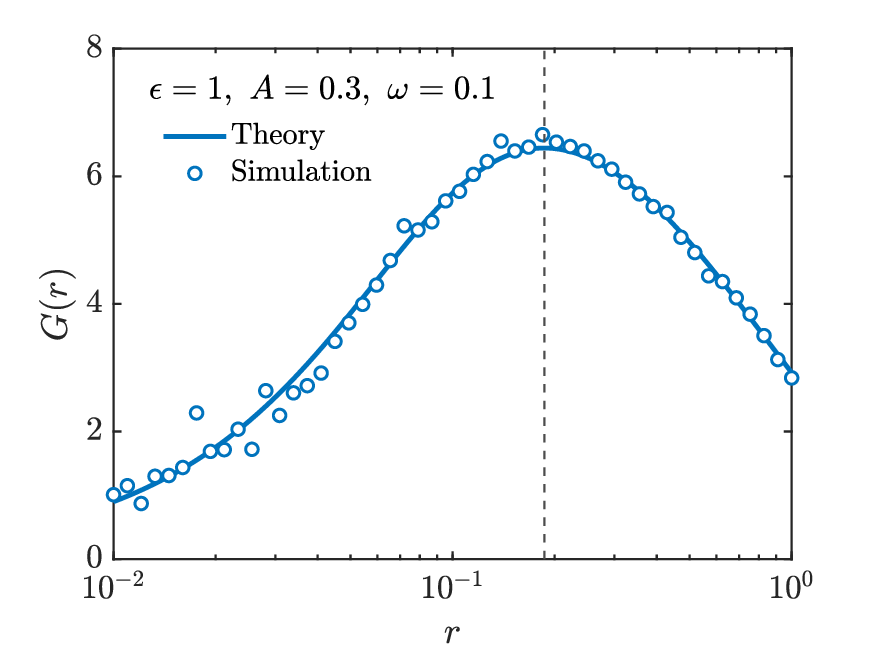}
(c)\includegraphics[width=0.9\linewidth]{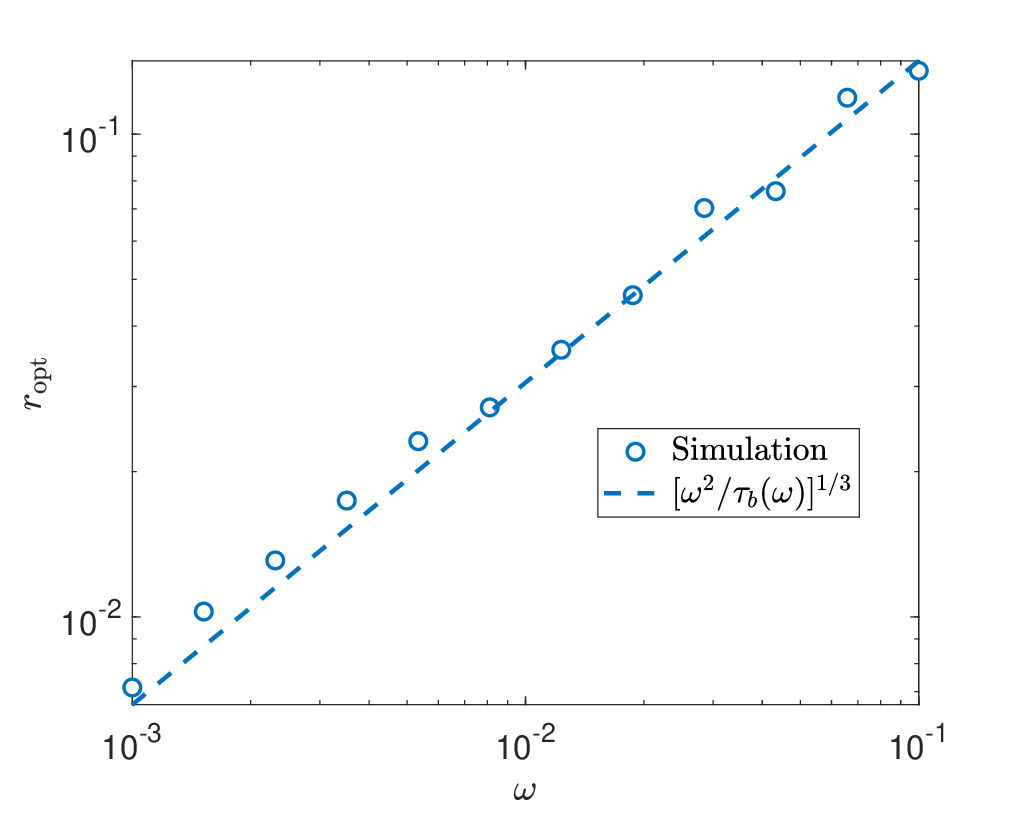}
\caption{
Three-stable-state realization of SRIR.
(a) Reset-induced phase dependence of the selected attractor. Different colors
denote three attracting branches.
(b) Amplification factor $G(r)$, showing a maximum at the finite optimal
SR rate $r_{\rm opt}\approx 0.18$ (dashed line). The solid line represents the prediction from the exact renewal formulation, while the open circles denote direct Monte Carlo simulations.
(c) Optimal SR rate versus driving frequency. Open circles denote
direct simulations and the dashed line shows the theoretical prediction
(\ref{sm:gen:optimal_scaling}).
Parameters are $\epsilon=1$ and $A=0.3$; panels (a) and (b) use
$\omega=0.1$.
}
\label{fig:SM}
\end{figure}

To demonstrate that the phase-selected finite-relaxation structure is not restricted to a bistable or one-dimensional geometry, we consider a two-dimensional system with three symmetry-related attracting states. In polar coordinates, the unforced potential is \begin{equation} U(\rho,\vartheta) = \frac{\rho^4}{4} - \frac{\rho^2}{2} - \epsilon\rho^3\cos(3\vartheta), \label{sm:three_U} \end{equation} and the periodically driven dynamics is \begin{equation}
\frac{d\mathbf x}{dt}
= -\nabla U(\mathbf x) + A \begin{pmatrix} \cos(\omega t)\\ \sin(\omega t) \end{pmatrix}, \label{sm:three_dynamics} \end{equation} with Poisson resetting to the origin. In Cartesian coordinates, one has
\begin{align} 
dx/dt ={}& x-(x^2+y^2)x + 3\epsilon(x^2-y^2) + A\cos(\omega t), \nonumber\\ dy/dt ={}& y-(x^2+y^2)y - 6\epsilon xy + A\sin(\omega t). \label{sm:three_cartesian} \end{align} In the absence of forcing, the three attractors (A) lie at \begin{equation} \vartheta_j = \frac{2\pi(j-1)}{3}, \qquad j=\{1,2,3\}, \label{sm:three_angles} \end{equation} with radius \begin{equation} \rho_A = \frac{ 3\epsilon+\sqrt{9\epsilon^2+4} }{2}. \label{sm:three_radius} \end{equation} They are separated by three saddle (S) at $\vartheta=(2j-1)\pi/3$ with radius \begin{equation} \rho_S = \frac{ -3\epsilon+\sqrt{9\epsilon^2+4} }{2}. \end{equation} The origin is a repelling point of the unforced system, see Fig.~1 in the main text. We choose the forcing sufficiently weak that, after the initial post-reset branch selection, the deterministic trajectory remains associated with the selected attracting branch until another reset occurs. Let $j(\phi)\in\{1,2,3\}$ denote the branch selected by an uninterrupted trajectory initiated at the origin at forcing phase $\phi$. Introducing  \begin{equation} \tau_s(\phi) = \inf\{ \tau>0: \rho_\phi(\tau)=\rho_A \}, \label{sm:three_tau} \end{equation} and assign the complex selected output \begin{equation} Q_{\rm sel}(\phi) = \rho_A \exp\left( \frac{2\pi i}{3} \bigl(j(\phi)-1\bigr) \right). \label{sm:three_Qsel} \end{equation} Equations~(\ref{sm:three_tau}) and (\ref{sm:three_Qsel}) have exactly the structure assumed in Eq.~(\ref{eq:general_step_response}).  These quantities enter directly into the
moments (\ref{sm:gen:moments}) in Sec.~\ref{sm-sec-2},  giving $\tau_b=\operatorname{Re}\left(M_1/M_0\right)$.
For the parameters considered in Fig.~\ref{fig:SM},
$\tau_b>0$ and the second absolute weighted moment remains
finite, so that the conditions of the phase-selected reduction are
satisfied.

Figure~\ref{fig:SM}(a) shows the phase-dependent
relaxation toward the three attracting branches.
Each reset phase selects one of the three attracting branches.
The Poisson SR dynamics also exhibits a pronounced maximum of
the coherent response at a finite resetting rate, as shown in
Fig.~\ref{fig:SM}(b). 
Figure~\ref{fig:SM}(c) compares the numerically determined
$r_{\rm opt}$ with the theoretical prediction (\ref{sm:gen:optimal_scaling}).
The numerical results follow this predicted scaling $r_{\rm opt}(\omega)
\propto
\omega^{2/3}$ over the
frequency range considered. 

More three- and four-stable-state rotational systems considered in
Ref.~\cite{zhu2024} also exhibit the same optimal-rate scaling
although these results are not shown here. These findings demonstrate that the competition between repeated resetting and finite post-reset relaxation extends beyond bistability.

\section{Selective control of SRIR and NIR}
\label{sm-sec-3}

SRIR and conventional NIR both exploit a periodically induced symmetry
breaking, but the driving acts on different dynamical objects. In
conventional NIR, it modulates noise-activated escape from the wells, whereas
in SRIR it biases the outgoing motion after reinjection near an unstable
state. We distinguish these two coupling mechanism and construct here the examples
in which one is present while the other is absent. 

Consider a symmetric bistable potential $U_0(x)$ with stable states
$x_\pm$ and an intermediate unstable state $x_u$, satisfying

\begin{eqnarray}
U_0'(x_\pm)&=&U_0'(x_u)=0,\nonumber\\
U_0''(x_\pm)&>&0,\quad U_0''(x_u)<0,\label{eq:symmetry_fixed_points}
\end{eqnarray} where the prime and double prime denote the first and second derivatives with respect to $x$, respectively.
Introduce a weak periodic perturbation through
\begin{eqnarray}
	U(x,t)&=&U_0(x)-A g(x)\cos(\omega t),	\nonumber\\
	d x/dt&=&-U_0'(x)+A g'(x)\cos(\omega t).\label{eq:general_periodic_dynamics}
\end{eqnarray}

The main equation (1) of the main text considered corresponds to
$g(x)=x$. In the following, We compare noise-free dynamics with Poisson resetting to
$x_u$ against dynamics without resetting, but with additive Gaussian
white noise of intensity $D$.

\subsection{Distinct symmetry-breaking mechanisms}

The different physical origins of the two resonances can be made
explicit for the general periodic coupling introduced above. We first
consider the dynamics immediately after resetting to the unstable state
$x_u$. Writing $x=x_u+\delta x$, the initial post-reset motion in the
quasistatic regime obeys, to leading order,
\begin{eqnarray}
	\frac{d(\delta x)}{dt}&\sim&\lambda_u\delta x
	+A g'(x_u)\cos\phi,
	\label{eq:local_SRIR_dynamics}
\end{eqnarray}
where $\lambda_u=-U_0''(x_u)>0$ and $\phi$ is the drive phase at resetting.
The phase-dependent bias of the outgoing motion is therefore controlled
by the parameter
\begin{eqnarray}
	\chi_{\rm SRIR}&\equiv&g'(x_u).
	\label{eq:chi_SRIR}
\end{eqnarray}
For $\chi_{\rm SRIR}\neq0$, the periodic signal biases the direction in
which the system leaves the unstable reset state. Provided that the
outgoing branches lead to distinct attracting basins, this local bias
selects the subsequent attractor and is amplified by the instability.
A finite optimal resetting rate additionally requires the competition
between phase selection and interruption of response build-up discussed 
in the main text. If instead $g'(x_u)=0$
the leading signal-induced bias at the reset state vanishes.

Conventional stochastic resonance probes a different aspect of the
same periodic coupling. Let $x_u(t)$ and $x_\pm(t)$ denote the
instantaneous saddle and local minima, which approach $x_u$ and
$x_\pm$ as $A\to0$. For weak forcing,
\begin{eqnarray}
	x_i(t)&=&x_i+\delta x_i(t),
	\qquad \delta x_i(t)=O(A),
\end{eqnarray}
where $i=\{u,+,-\}$ and $U_0'(x_i)=0$. Expanding the potential at the
shifted extrema gives
\begin{eqnarray}
	U[x_i(t),t]
	&=&
	U_0[x_i+\delta x_i(t)]
	-
	A g[x_i+\delta x_i(t)]\cos(\omega t)
	\nonumber\\
	&=&
	U_0(x_i)+U_0'(x_i)\delta x_i(t)
	\nonumber\\
	&&{}
	-A\left[g(x_i)+g'(x_i)\delta x_i(t)\right]
	\cos(\omega t)
	+O(A^2)
	\nonumber\\
	&=&
	U_0(x_i)-A g(x_i)\cos(\omega t)
	+O(A^2),\nonumber\\
\label{eq:potential_extrema_first_order}
\end{eqnarray} where we used $U_0'(x_i)=0$ and $\delta x_i(t)=O(A)$, so that
$A g'(x_i)\delta x_i(t)=O(A^2)$.
The instantaneous activation barriers are
therefore
\begin{eqnarray}
	\Delta U_\pm(t)
	&=&
	U[x_u(t),t]-U[x_\pm(t),t]
	\nonumber\\
	&=&
	U_0(x_u)-U_0(x_\pm)
	\nonumber\\
	&&-
	A[g(x_u)-g(x_\pm)]\cos(\omega t)
	+O(A^2).\nonumber\\
\label{eq:barrier_general}
\end{eqnarray}

The directional
barrier asymmetry is
\begin{eqnarray}
	\Delta U_+(t)-\Delta U_-(t)
	&=&A\chi_{\rm NIR}\cos(\omega t)+O(A^2),\nonumber\\
\end{eqnarray} 
where \begin{eqnarray}
\chi_{\rm NIR}&\equiv&g(x_+)-g(x_-).
\label{eq:barrier_asymmetry}
\end{eqnarray}
Thus $\chi_{\rm NIR}\neq0$ values periodically favor noise-induced escape from
one well over the other and provides the conventional route to coherent
interwell switching. When $g(x_+)=g(x_-),$
the leading directional barrier modulation vanishes. This does not
imply that the complete noise-driven response vanishes, because common
barrier modulation, curvature-dependent prefactors, intrawell motion,
and higher-order barrier changes may still contribute.

The two resonance mechanisms therefore probe different properties of
periodic coupling, meansured by the quantitis (\ref{eq:chi_SRIR}) and (\ref{eq:barrier_asymmetry}) respectively.
The former is determined locally by  forcing near the unstable reset
state, whereas the latter reflects the global energetic asymmetry
between the two wells. These quantities can be varied independently,
allowing the post-reset phase-selection mechanism to remain active when
the leading barrier-asymmetry mechanism is suppressed, or \textit{vice versa}.

For the standard additive forcing as discussed in the main text, $g(x)=x$, one has
\begin{eqnarray}
	\chi_{\rm SRIR}&=&1,\quad
	\chi_{\rm NIR}=x_+-x_-,
	\label{eq:standard_g}
\end{eqnarray}
so that both mechanisms are present. Thus, in the standard periodically
tilted bistable system, the same forcing can simultaneously bias
noise-induced barrier crossing and post-reset attractor selection.

The examples below realize two different cases:
the first retains the post-reset phase-selection mechanism while
suppressing the leading barrier asymmetry, i.e.,
\begin{eqnarray}
\chi_{\rm SRIR}\neq0, \quad \chi_{\rm NIR}=0
\end{eqnarray} whereas the second retains
the leading barrier-asymmetry mechanism while suppressing post-reset
phase selection, so that
\begin{eqnarray}
\chi_{\rm SRIR}=0, \quad \chi_{\rm NIR}\neq0.
\end{eqnarray}

\subsection{Example 1: Active SRIR and suppressed NIR}

Consider the quartic double well
\begin{eqnarray}
 U_0(x)&=&-\frac{a}{2}x^2+\frac{b}{4}x^4,
 \label{eq:quartic_for_symmetry}
\end{eqnarray} where $ x_u=0, x_\pm=\pm x_s,
x_s=\sqrt{a/b}.$
We choose first
\begin{eqnarray}
 g(x)&=&x\left(1-\frac{x^2}{x_s^2}\right).
 \label{eq:g_SRIR_only}
\end{eqnarray} and   
the driven deterministic equation becomes
\begin{eqnarray}
	\frac{dx}{dt}&=&ax-bx^3
	+A\left(1-\frac{3x^2}{x_s^2}\right)\cos(\omega t).
	\label{eq:SRIR_only_dynamics}
\end{eqnarray}
Then we have
\begin{eqnarray}
 \chi_{\rm SRIR}&=&1,\qquad\chi_{\rm NIR}=0.
 \label{eq:g_R_channels}
\end{eqnarray}
Near $x=0$, its quasistatic form is
$\dot x\sim ax+A\cos\phi.$
Hence the same leading local seed as in the main-text model is
retained. Meanwhile, for sufficiently weak forcing, $\Delta U_\pm(t)=\Delta U_0+O(A^2).$
Both activation barriers have no first-order height modulation because
$g(x)$ in Eq.~(\ref{eq:g_SRIR_only}) vanishes at all three unperturbed extrema. This construction
therefore preserves the local SRIR seeding channel while removing the
leading barrier-height modulation.

\begin{figure}
\centering
(a)\includegraphics[width=0.9\linewidth]{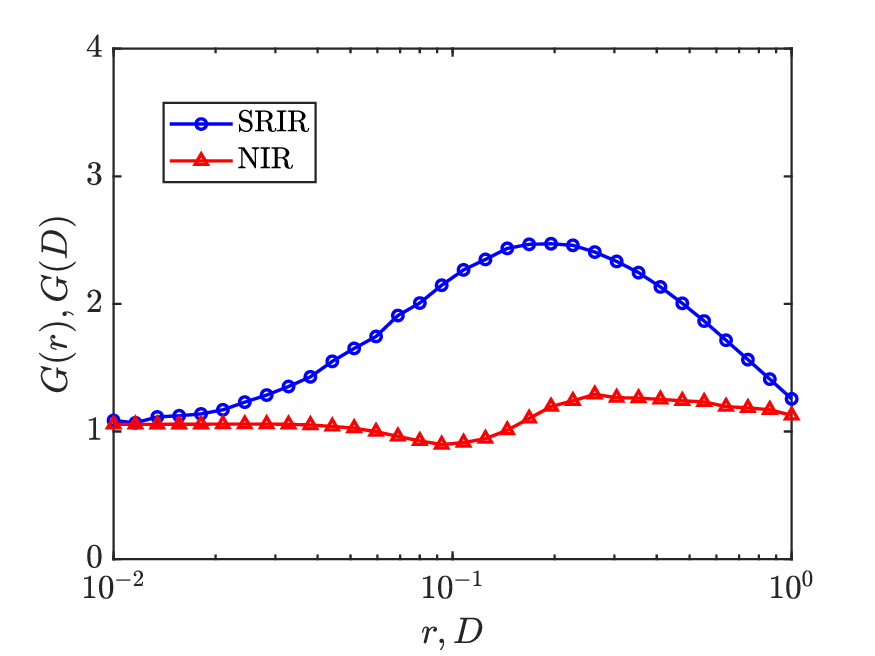}
(b)\includegraphics[width=0.9\linewidth]{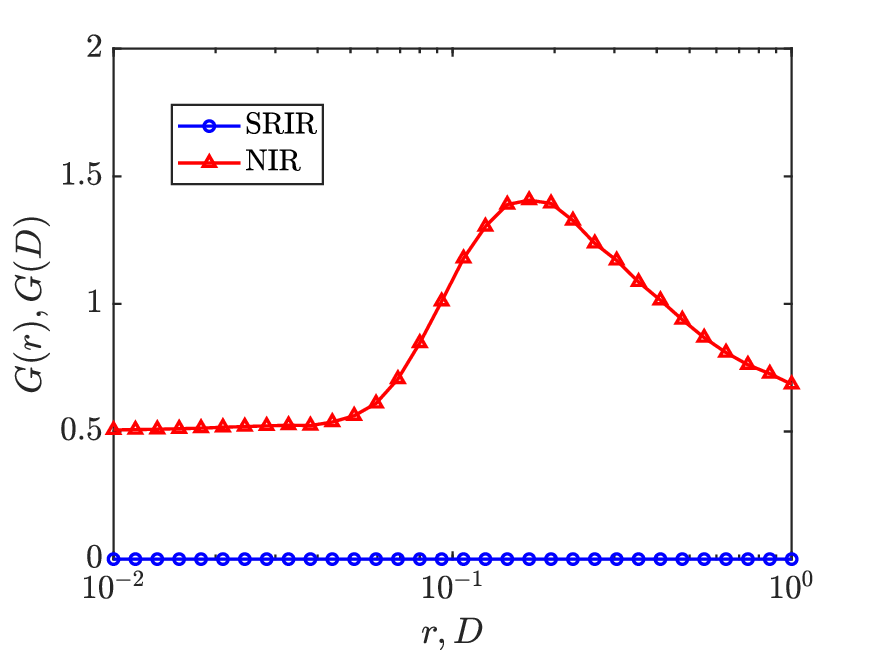}
\caption{
Comparison of SRIR and NIR for two nonlinear systems (\ref{eq:general_periodic_dynamics}).
(a) For $g(x)$ given by Eq.~(\ref{eq:g_SRIR_only}), SRIR (blue circles) exhibits a pronounced
maximum as a function of SR rate $r$, while NIR
(red triangles) is much weaker.
(b) For $g(x)$ given by Eq.~(\ref{eq:g_SR_only}), NIR displays a clear maximum as a function
of the noise intensity $D$, whereas the SRIR response vanishes
identically, $G=0$, because resetting generates no phase-dependent bias.
The two examples demonstrate that SRIR and NIR are dynamically distinct
and that the existence of one does not imply the other. Parameters: $A=0.3,\omega=0.1,a=b=x_s=1$.
}
\label{fig:comparison}
\end{figure}

Figure~\ref{fig:comparison}(a) shows a case in which SR produces a pronounced resonant enhancement. The SRIR gain
develops a clear maximum at an intermediate resetting rate, whereas
the corresponding NIR response remains much weaker over the same
parameter range. This contrast indicates that the strong response is
not a generic consequence of stochastic fluctuations. Rather, it
originates from the reset-induced phase selection, through which
reinitialization repeatedly places the system in a dynamically
susceptible configuration where the weak periodic bias can efficiently
select and amplify the subsequent response.

\subsection{Example 2: Suppressed SRIR and active NIR}

For the same quartic potential, we choose now instead
\begin{eqnarray}
 g(x)&=&\frac{x^3}{3x_s^2}.
 \label{eq:g_SR_only}
\end{eqnarray}
The deterministic dynamics is
\begin{eqnarray}
	\frac{dx}{dt}&=&ax-bx^3
	+A\frac{x^2}{x_s^2}\cos(\omega t).
	\label{eq:SR_only_dynamics}
\end{eqnarray} Then we have
\begin{eqnarray}
 \chi_{\rm SRIR}&=&0,\qquad\chi_{\rm NIR}=\frac{2x_s}{3}.
 \label{eq:g_S_channels}
\end{eqnarray}

At the ideal reset point, $\dot x=0,$
for every driving phase. Thus a trajectory reset exactly to $x_r=0$
remains there for all subsequent times in the noise-free dynamics.
The deterministic phase-selection mechanism is absent, not merely
small to first order. Nevertheless, the activation barriers obey $
 \Delta U_+(t)-\Delta U_-(t)
 =\frac{2A x_s}{3}\cos(\omega t)+O(A^2).
$
The conventional directional barrier-modulation channel therefore
remains present, and noise-assisted switching can respond to the driving.
Whether the chosen response measure has a pronounced maximum as $D$
is varied must be checked at the specified forcing and frequency.
Finite reset-position uncertainty or additional noise removes the
exact trapping at $x=0$ and can permit a reset response not captured by
the ideal seeding criterion.

The complementary behavior is illustrated in
Fig.~\ref{fig:comparison}(b). Here the noise-driven system exhibits
a clear resonant maximum at a finite noise intensity, while the
resetting-induced coherent response vanishes identically for all SR rates. In this case the prescribed
reset state carries no phase-dependent bias, so resetting does not
select a coherent outgoing response even though the underlying
nonlinear system supports NIR. This example shows that SR alone is insufficient to generate resonance. A
phase-sensitive post-reset response is an essential ingredient of SRIR.

\section{Numerical methods}
\label{sm-sec-4}

\subsection{Direct Monte Carlo simulation}
\label{subsec:direct_simulation}

Between successive resets, the noise-free bistable dynamics obeys
\begin{equation}
\frac{dx}{dt} =ax-bx^3+A\cos(\omega t).
\label{eq:num_reset_dynamics}
\end{equation}
At each Poisson reset time $t_n$, the position is returned to
$x(t_n)=0$, while the external drive phase remains unchanged.

The probability of at least one SR event during a numerical step
$\Delta t$ is
\begin{equation}
p_r=1-e^{-r\Delta t}.
\label{eq:num_reset_probability}
\end{equation}
Equivalently, continuous Poisson event times can be generated from
$t_{n+1}-t_n=-\ln\left( U_n/r\right)$, where the $U_n$ are independent uniform
random variables on $(0,1)$. 


The response is measured over
an integer number of driving periods,
$T_{\rm obs}=N_{\rm per}T$, where $T=2\pi/\omega$.
For realization $j$, we define
\begin{align}
Q_c^{(j)}
&=
\frac{2}{T_{\rm obs}}
\int_{t_0}^{t_0+T_{\rm obs}}
x_j(t)\cos(\omega t)\,dt,
\nonumber\\
Q_s^{(j)}
&=
\frac{2}{T_{\rm obs}}
\int_{t_0}^{t_0+T_{\rm obs}}
x_j(t)\sin(\omega t)\,dt.
\label{eq:num_Qcs}
\end{align}
For $N_{\rm ens}$ independent realizations, the ensemble-averaged
components are
$\overline Q_\alpha=N_{\rm ens}^{-1}
\sum_{j=1}^{N_{\rm ens}}Q_\alpha^{(j)}$,
with $\alpha=\{c,s\}$. Thus, we set
\begin{equation}
C_{1}
=
\frac{\overline Q_c-i\overline Q_s}{2},
\qquad
G(r)
=
\frac{
\sqrt{\overline Q_c^{\,2}+\overline Q_s^{\,2}}
}{A}.
\label{eq:num_Gsim}
\end{equation}
The Fourier components are therefore ensemble averaged before their
magnitude is taken.

For the three-stable-state system (\ref{sm:three_dynamics}),  the periodic response is extracted from the
complex position $Z_j(t)=x_j(t)+iy_j(t)$.
For each realization, we calculate the
Fourier coefficient
\begin{equation}
C_{1}^{(j)}
=
\frac{1}{T_{\rm obs}}
\int_{t_0}^{t_0+T_{\rm obs}}
Z_j(t)e^{-i\omega t}\,dt.
\end{equation}
The ensemble-averaged coefficient and the
corresponding amplification factor are
\begin{eqnarray}
C_{1}
=
{N^{-1}_{\rm ens}}
\sum_{j=1}^{N_{\rm ens}}C_{1}^{(j)},\quad
G(r)=
\frac{|C_{1}|}{A}.
\label{eq:MC_complex_gain}
\end{eqnarray}

\subsection{Post-reset trajectories}
\label{subsec:num_phase_trajectories}

The renewal and coarse-grained reduction
use the same uninterrupted post-reset trajectories. The reset phase is
sampled on the midpoint grid
\begin{equation}
\phi_j
=
\frac{2\pi(j+1/2)}{N_\phi},
\qquad
j=\{0,\ldots,N_\phi-1\}.
\label{eq:num_phase_grid}
\end{equation}
For each $\phi_j$, the trajectory is obtained by integrating
\begin{subequations}\label{eq:num_post_reset_dynamics}
\begin{align}
\frac{d x_{\phi_j}}{d\tau}
&=
a x_{\phi_j}-b x_{\phi_j}^3
+A\cos(\phi_j+\omega\tau),\\
x_{\phi_j}(0)&=0,
\end{align}    
\end{subequations}
with a fourth-order Runge--Kutta scheme.
Here $\tau$ is the elapsed time since the reset, and no further reset
is applied while computing $x_{\phi_j}(\tau)$.

The first phase harmonic is
\begin{align}
h_1(\tau)
&=
\frac{1}{2\pi}
\int_0^{2\pi}
x_\phi(\tau)e^{-i\phi}\,d\phi
\nonumber\\
&\sim
\frac{1}{N_\phi}
\sum_{j=0}^{N_\phi-1}
x_{\phi_j}(\tau)e^{-i\phi_j}.
\label{eq:num_h1}
\end{align}
The exact renewal representation then gives
\begin{align}
C_{1}(r)
&=
r\int_0^\infty
e^{-(r+i\omega)\tau}h_1(\tau)\,d\tau,
\nonumber\\
G(r)
&=
\frac{2|C_{1}(r)|}{A}.
\label{eq:num_renewal_response}
\end{align}
For numerical evaluation, the age integral is truncated at
$\tau_{\max}$ and discretized as
\begin{equation}
C_{1}(r)
\sim
r\sum_{k=0}^{N_\tau}
w_k e^{-(r+i\omega)\tau_k}h_1(\tau_k),
\label{eq:num_renewal_quadrature}
\end{equation}
where $\tau_k=k\Delta\tau$,
$\tau_{\max}=N_\tau\Delta\tau$, and $w_k$ are the quadrature weights. 

For the bistable realization, the complete trajectory is
replaced by
\begin{equation}
x_\phi(\tau)
=
\sigma_\phi x_s\Theta[\tau-\tau_s(\phi)],
\qquad
x_s=\sqrt{a/b},
\label{eq:num_step_trajectory}
\end{equation}
where
\begin{align}
\tau_s(\phi)
&=
\inf\{\tau>0:|x_\phi(\tau)|=x_s\},
\nonumber\\
\sigma_\phi
&=
\operatorname{sgn}[x_\phi(\tau_s(\phi))].
\label{eq:num_arrival_definition}
\end{align}
If the first threshold crossing occurs between $\tau_k$ and
$\tau_{k+1}$, its time is refined by linear interpolation
\begin{equation}
\tau_s(\phi_j)
\sim
\tau_k+\Delta\tau
\frac{
\sigma_{\phi_j}x_s-x_{\phi_j}(\tau_k)
}{
x_{\phi_j}(\tau_{k+1})-x_{\phi_j}(\tau_k)
},
\label{eq:num_crossing_interpolation}
\end{equation}
with
$\sigma_{\phi_j}=\operatorname{sgn}[x_{\phi_j}(\tau_{k+1})]$.

Substitution into the renewal formula yields
\begin{equation}
C_1(r)
=
\frac{x_s r}{2\pi(r+i\omega)}
\int_0^{2\pi}
\sigma_\phi
e^{-i\phi-(r+i\omega)\tau_s(\phi)}\,d\phi.
\label{eq:num_step_C1}
\end{equation}
Consequently, the phase-grid approximation to the reduced gain is
\begin{equation}
G(r)
\sim
\frac{2x_s r}{A\sqrt{r^2+\omega^2}}
\left|
\frac{1}{N_\phi}
\sum_{j=0}^{N_\phi-1}
\sigma_{\phi_j}
e^{-i\phi_j-(r+i\omega)\tau_s(\phi_j)}
\right|.
\label{eq:num_step_gain}
\end{equation}
The SR rate is varied without recomputing the uninterrupted
trajectories.
Scanning and maximization give the optimal rates for
the renewal and give rise to reduced response.


For the three-stable-state system (\ref{sm:three_dynamics}), the two-dimensional
dynamics can be expressed in terms of the complex
position $Z=x+iy$ as
\begin{equation}
\frac{dZ}{dt}=Z-|Z|^2Z+3\epsilon\overline Z^{\,2}
+Ae^{i\omega t},
\end{equation}
with Poisson SR to $Z=0$.

The renewal formulation applies directly to the
complex post-reset trajectories $Z_\phi(\tau)$.
The first-harmonic coefficient and amplification
factor are given by
\begin{subequations}\label{eq:complex_gain}
 \begin{eqnarray}
h_{1}(\tau)
&=&\frac{1}{2\pi}\int_0^{2\pi}
Z_\phi(\tau)e^{-i\phi}\,d\phi,\\
C_{1}(r)
&=&r\int_0^\infty
e^{-(r+i\omega)\tau}h_{1}(\tau)\,d\tau,\\
G(r)
&=&\frac{|C_{1}(r)|}{A}.
\end{eqnarray}   
\end{subequations}

\subsection{Time-Periodic Fokker--Planck Calculations for NIR}
\label{sec:FPK_noise_gain}

For comparison with conventional NIR, resetting is
removed and additive Gaussian white noise is included
\cite{Gammaitoni1998}
\begin{equation}
\frac{dx}{dt}
=
ax-bx^3+A\cos(\omega t)+\sqrt{2D}\,\xi(t),
\label{eq:noise_langevin}
\end{equation}
where $\langle\xi(t)\rangle=0$ and
$\langle\xi(t)\xi(t')\rangle=\delta(t-t')$.
The probability density satisfies \cite{Risken1989}
\begin{align}
\frac{\partial P(x,t)}{\partial t}
={}&
-\frac{\partial \bigl[(ax-bx^3+A\cos\omega t)P(x,t)\bigr]}{\partial t}\nonumber\\&
+D\frac{\partial^2 P(x,t)}{\partial x^2}
.
\label{eq:FP_noise}
\end{align}
The periodic density is approximated by the truncated harmonic
expansion
\begin{equation}
P(x,t)
\sim
\sum_{n=-N_h}^{N_h}p_n(x)e^{in\omega t},
\label{eq:FP_fourier}
\end{equation}
Here, $p_{-n}(x)=p_n^*(x)$, $N_h$ denotes the Fourier truncation order corresponding to the highest harmonic retained in the expansion, and the superscript $(*)$ denotes complex conjugation.

Substituting Eq.~(\ref{eq:FP_fourier}) into Eq.~(\ref{eq:FP_noise}) one can yield
\begin{equation}
in\omega p_n
=-\frac{\partial J_n}{\partial_x}
\label{eq:FP_harmonic_balance}
\end{equation}
where
\begin{eqnarray}
J_n
=
(ax-bx^3)p_n
+
\frac{A}{2}(p_{n-1}+p_{n+1})
-
D\frac{\partial p_n}{\partial_x}.\nonumber\\
\label{eq:FP_harmonic_current}
\end{eqnarray}
The harmonic truncation is given by
$p_{N_h+1}=p_{-N_h-1}=0$.

On the computational domain $[x_L,x_R]$, 
zero-flux boundary conditions and zero-mode normalization are imposed
via
\begin{eqnarray}
J_n(x_L)=J_n(x_R)&=&0,
\nonumber\\
\int_{x_L}^{x_R}p_0(x)\,dx&=&1.
\label{eq:FP_boundary_normalization}
\end{eqnarray}
The second condition obeys as for $n\neq0$, integrating (\ref{eq:FP_harmonic_balance})
over the domain $[x_L,x_R]$ together with the zero-flux
boundary conditions gives
$\int_{x_L}^{x_R}p_n(x)\,dx=0$.
Thus, only the zero-frequency mode carries the total probability.

The spatial discretization employs a conservative
finite-volume scheme. We divide the computational
domain $[x_L,x_R]$ into spatial cells
$[x_{i-1/2},x_{i+1/2}]$ of width
$\Delta x_i=x_{i+1/2}-x_{i-1/2}$.
The cell-averaged Fourier coefficient is defined as
\begin{equation}
\overline p_{n,i}
=
\frac{1}{\Delta x_i}
\int_{x_{i-1/2}}^{x_{i+1/2}}
p_n(x)\,dx.
\end{equation}
Integrating the harmonic conservation equation
(\ref{eq:FP_harmonic_balance}) over each cell yields
\begin{equation}
in\omega\overline p_{n,i}
=
-\frac{
J_{n,i+1/2}-J_{n,i-1/2}
}{\Delta x_i},
\label{eq:FP_finite_volume}
\end{equation}
where $J_{n,i\pm1/2}$ denote the numerical probability
fluxes through the cell interfaces, approximating
the continuous harmonic current
$J_n(x_{i\pm1/2})$ defined in
Eq.~(\ref{eq:FP_harmonic_current}).
The same interface flux is shared by adjacent cells,
so that the internal flux contributions cancel
when the discrete equations are summed over
the computational domain.
The coupled harmonic equations (\ref{eq:FP_finite_volume}) form a linear system for the cell-averaged
Fourier coefficients ${\overline p_{n,i}}$.
Solving this system under zero-flux boundary conditions $J_{n,1/2}=J_{n,N_x+1/2}=0$ and the normalization constraint $\sum_i\overline p_{0,i}\Delta x_i=1$ yields the Fourier coefficients of the time-periodic stationary probability density. 

The first-harmonic coefficient of the mean response
is obtained from
\begin{equation}
C_{1,\rm noise}
=
\int_{-\infty}^{\infty}x\,p_1(x)\,dx
\sim
\sum_i x_i\overline p_{1,i}\Delta x_i,
\label{sm:int:noise_C1}
\end{equation}
where $x_i$ denotes the cell center and the
numerical sum extends over the computational domain.
The corresponding amplification factor is
\begin{equation}
G_{\rm noise}(D;A,\omega)
=
\frac{2|C_{1,\rm noise}|}{A}.
\label{eq:G_noise_FP}
\end{equation}
For fixed driving amplitude $A$ and frequency
$\omega$, the maximum amplification is obtained
by optimizing over the noise intensity $D$,
\begin{equation}
G_{\rm noise}^{\max}(A,\omega)
=
\max_{D>0}G_{\rm noise}(D;A,\omega).
\label{sm:int:optimized_comparison}
\end{equation}
This approach directly solves the time-periodic
Fokker--Planck equation without invoking a
Kramers-rate or two-state approximation. This provide the results of NIR in Fig.~4 of the main text.

\setcounter{secnumdepth}{2}
\renewcommand{\thesection}{S\arabic{section}}
\renewcommand{\theequation}{S\arabic{equation}}

\section{Dynamical Visualization of SRIR}
\label{sec:movies}

Supplemental Movies S1 and S2 illustrate the SRIR in the bistable and tristable systems discussed in
Sec.~S2.2 and Sec.~S2.3, respectively. 

\subsection{Supplemental Movie S1: Bistable system}

The upper panel displays the periodically driven
double-well potential and particle positions, while
the middle panel compares an individual trajectory
with the ensemble-averaged response. The lower panel
shows the amplification factor $G(r)$ as a function
of the resetting rate.

In Movie S1, particles are reset to the origin
($x=0$), an unstable state of the
undriven bistable system. For rare SR ($r=0.01$), the individual particle
(white sphere) remains predominantly trapped in one of
the stable wells, while the ensemble-averaged position
(yellow sphere) exhibits only weak periodic oscillations.
At the optimal SR rate ($r\approx 0.178$), the individual
particle undergoes irregular transitions between the
two wells following SR, whereas
the ensemble-averaged position exhibits pronounced,
resonant oscillations.
For frequent SR ($r=1$), repeated returns
to the unstable point interrupt relaxation toward
the stable wells, confining the individual particle
closer to the origin and strongly suppressing the
ensemble-averaged oscillations.

\subsection{Supplemental Movie S2: Tristable system}
The upper panel presents a three-dimensional rendering of
the time-dependent potential landscape, illustrating the
three stable attractors and the evolution of particle
positions, with different colors distinguishing the
three attractors.  The lower-left panels display the
$x$ and $y$ components of an individual trajectory
and the ensemble-averaged response, while the
lower-right panel shows the corresponding
amplification factor $G(r)$.

In Movie S2, particles are reset to the origin
($x=y=0$), an unstable state of the
undriven system. For rare SR ($r=0.01$), the individual particle
(white sphere) spends long periods close to a single
attractor, whereas the ensemble-averaged position
(yellow sphere) shows only a weak response to the
rotating periodic drive.
At the optimal SR rate ($r\approx 0.185$), the individual
particle is randomly redirected among the three
attractors through SR, while the ensemble
average develops periodic large-scale oscillations
in both spatial components, following the rotating
drive.
At high SR rates ($r=1$), frequent
reinitialization prevents the particle from fully
approaching the selected attractors, resulting in
a substantially reduced ensemble response.
The movie demonstrates that SRIR extends beyond
bistability, producing resonant two-dimensional
dynamics even though individual trajectories
remain stochastic.